%% file: main.tex
\documentclass{article}
\usepackage{paper}
\usepackage{subcaption}
\newcounter{num}
\makeatletter

\newcommand{\TODO}[1]{\textcolor{red}{[TODO\@ifnotempty{#1}{: #1}]}}
\newcommand{\fred}[1]{\textcolor{purple}{[fred\@ifnotempty{#1}{: #1}]}}
\newcommand{\sandeep}[1]{\textcolor{orange}{[sandeep\@ifnotempty{#1}{: #1}]}}
\newcommand{\ali}[1]{\textcolor{blue}{[ali\@ifnotempty{#1}{: #1}]}}
\makeatother

\usepackage{algorithm}
\usepackage{algpseudocode}
\usepackage{wrapfig}

\title{Faster Fundamental Graph Algorithms via Learned Predictions}
\author{Justin Y.\ Chen \\ MIT \\ \texttt{justc@mit.edu}  \and Sandeep Silwal\\ MIT  \\ \texttt{silwal@mit.edu}\and Ali Vakilian\\ TTIC\\ \texttt{vakilian@ttic.edu}  \and Fred Zhang \\ UC Berkeley  \\
  \texttt{z0@berkeley.edu}}
\date{}

\begin{document}

\maketitle
\begin{abstract}
We consider the question of speeding up classic graph algorithms with machine-learned predictions.
In this model, algorithms are furnished with extra advice learned from past or similar instances. Given the additional information, we aim to improve upon the traditional worst-case run-time guarantees.   Our contributions are the following:
\begin{enumerate}[(i)]
    \item We give a faster algorithm for minimum-weight bipartite matching via learned duals, improving   the recent result by Dinitz, Im, Lavastida, Moseley and Vassilvitskii (NeurIPS, 2021);
    \item We extend the learned dual approach to the single-source shortest path problem (with negative edge lengths), achieving an almost linear runtime given sufficiently accurate predictions which improves upon the classic fastest algorithm due to Goldberg (SIAM J. Comput., 1995);
    \item We provide a general reduction-based framework for learning-based graph algorithms, leading to new algorithms for degree-constrained subgraph and minimum-cost $0$-$1$  flow, based on reductions to bipartite matching and the shortest path problem.
\end{enumerate}
Finally, we give a set of general learnability theorems, showing that the predictions required by our algorithms can be efficiently learned in a PAC fashion.
\end{abstract}
\newpage
\input{intro}

\input{prelim} 

\input{matching}

\input{shorestpath_duals}

\input{reductions}

\input{learnable}

\input{experiment}

 \section*{Acknowledgment}
 We thank Piotr Indyk for helpful comments on an early draft of the paper and Robert Tarjan for providing us with the reference~\cite{gabow1983efficient}.
 
 Justin and Sandeep are supported by an NSF Graduate Research Fellowship under Grant No.\ 1745302, NSF TRIPODS program (award DMS-2022448), NSF award CCF-2006798, and Simons Investigator Award. Justin is also supported by a MathWorks Engineering Fellowship.
%Justin is supported by a MathWorks Engineering Fellowship and NSF Graduate Research Fellowship under Grant No.\ 1745302. Sandeep is supported by an NSF Graduate Research Fellowship under Grant No.\ 1745302.
 Ali is supported by NSF award CCF-1934843.

\bibliographystyle{alpha}
\bibliography{bib}
\newpage
\appendix
%\section{Learning-based Push-relabel Maximum Flow}
\input{icml22/other} 
\input{icml22/omit3}

\input{icml22/omit-bmatch}

\input{icml22/omit4}
\input{icml22/additional-reduction}

\input{icml22/omit6}

\end{document}

%% file: intro.tex
\section{Introduction}
There has been recent interest in moving beyond the traditional and often pessimistic worst-case analysis of algorithms by using machine-learned predictions. This paradigm of \textit{learning-augmented algorithms} is inspired by the great success of machine learning  (ML)   and aims to utilize  ML predictions to improve the performance of classic algorithms. 

The extra information assumed in learning-augmented algorithms can be supplied in a variety of settings. For example, in data streams, the observation that underlying patterns in real-world datasets do not change quickly over time has led to the development of an oracle capable of predicting frequently occurring stream elements. In distribution learning, it is natural to have access to different but related distributions that can aid our learning tasks. In many other applications, the current input can be similar to past instances that might help us to avoid computing the solution from scratch.  All these scenarios fall under the general umbrella of a ``warm start'', which enables better initialization of the algorithms to improve their performance.

This learning-based paradigm  has been successfully applied in many algorithmic domains. They all share an underlying goal to minimize some resource constraints: in online algorithms, predictions are used to make better future decisions and reduce regret and competitive ratios~\cite{lykouris2018competitive}. In streaming algorithms and data structures, predictors have been developed to optimize space usage~\cite{kraska2017case, hsu2018}. In sublinear algorithms, predictors can reduce the sample complexity of a task~\cite{EINR+2021}. %See Section \ref{sec:related_works} for a more comprehensive discussion on related works and other applications.

Despite this activity, only recently have there been works on provably improving the time complexity of algorithms under this framework.  Two recent works which consider this resource include the work of  \cite{ergun2021} on the $k$-means clustering problem and the work of \cite{dinitz2021faster} on graph matching; our paper relates to the latter. In \cite{dinitz2021faster}, the authors give a learning-based algorithm for (min-cost) bipartite matching and show that predictions provably result in faster algorithms. Our paper is motivated by three natural follow-up questions:
\begin{enumerate}[(i)]
    \item Can we derive learning-augmented algorithms which exploit warm starts and other auxilliary information for  other important graph optimization problems besides bipartite matching?
    \item Do we need a tailor made learning-augmented algorithm for every different graph optimization problem?
    \item Can we understand when warm starts are learnable for general problems?
\end{enumerate}

\subsection{Our Results}
%\sandeep{todo: add `informal' theorem statements and exact bounds we get}
Our main contributions provide answers to   the three motivating questions. We individually address these questions and our relevant contributions.
\begin{quote}
    \textit{Can we derive learning-augmented algorithms for other classic graph optimization problems besides bipartite matching?} 
\end{quote}
Towards answering this question, we first provide a more efficient learning-augmented algorithm for bipartite matching than the one in \cite{dinitz2021faster}, in \cref{sec:faster_matching}. The algorithm of \cite{dinitz2021faster} uses dual variables from the linear programming formulation of bipartite matching as predictions. We achieve better runtime by utilizing the interplay of this dual and another set of related dual variables, called reduced edge duals, arising from viewing bipartite matching as a max-flow problem.  This result  extends to $b$-matching as well.
\begin{theorem}[Informal; see \cref{thm:faster-matching}]
Given a weighted bipartite graph and predicted dual  $\hat{y}$ ,  there exists an algorithm that finds a minimum weight perfect matching in time $O(m\sqrt{n} + (m + n\log n) ||y^* - \hat{y}||_0)$, where $y^*$ is an optimal dual solution.
\end{theorem}
This significantly improves upon the prior bound of $\widetilde{O}(\min \{m \sqrt{n} ||y^* - \hat{y}'||_1, mn\})$ due to \cite{dinitz2021faster}.

Beyond the problem of minimum-weight matching, we also use reduced edge duals, which allow us to obtain the first learning-augmented algorithm for the single-source shortest-paths problem with negative edge lengths. 
\begin{theorem}[Informal; see \cref{thm:faster_shortest_paths_dual}]
Given a directed graph with negative edge weights and predicted dual  $\hat{y}$ , there exists an algorithm that  finds single-source shortest paths in time 
$O(m \min\{\|\hat{y}-y^*\|_1 \cdot \|\hat{y}-y^*\|_{\infty}, \sqrt{n} \log(\|\hat{y}-y^*\|_{\infty})\})$, where $y^*$ is an optimal dual solution.
\end{theorem}
To properly utilize these duals, we give an efficient rounding scheme which takes in as input a set of predicted reduced edge duals and rounds them to a feasible instance. See   \cref{sec:RE_dual} for the formal notion of feasibility and our rounding algorithm.
\begin{quote}
     \textit{Do we need a `tailor made' learning-augmented algorithm for every different graph optimization problem?} 
\end{quote}

The prior work \cite{dinitz2021faster} outlined the three challenges of ``feasibility, optimization, and learnabilty'' needed to put warm start heuristics on theoretical footing. We leverage \textit{reductions}   to avoid addressing these challenges from scratch for each new graph problem.   Specifically, we introduce a general framework of reductions that takes existing learning-augmented algorithms and applies them to new problems. Note that in the context of learning-augmented algorithms, we need reductions that efficiently convert instances of a given problem to instances of another problem which we know how to solve using predictions.  Therefore, we must judiciously choose the problems and reductions to apply in this framework. Nonetheless, the benefits of our reduction framework include faster learning-augmented algorithms for shortest-paths and new algorithms for other problems, such as degree-constrained subgraph and unit-capacity maximum flow.  
\begin{quote}
     \textit{Can we understand when warm starts are learnable for general graph problems?} 
\end{quote}
Given the wide range of problems we consider, we need to understand when good hints and predictions which generalize are learnable. (Note that \cite{dinitz2021faster} is only concerned about learnability of duals for the specific problem of bipartite matching.) We answer this question in   \cref{sec:learnability} by generalizing the arguments of \cite{dinitz2021faster} beyond bipartite matching. %Our proofs follow from bounding the pseudo-dimension of the class of loss functions that we are generally interested in learning LP duals on graphs.

\subsection{Related Work}\label{sec:related_works}
\paragraph{Learning-augmented graph algorithms.} The most relevant work to ours is~\cite{dinitz2021faster}. We improve and extend their results in several ways, as discussed earlier. For the shortest-path problem, a recent work \cite{eden2022embeddings} investigates the theory of learning-based labeling scheme for the $A^*$ search heuristic, whereas a few others approach it empirically \cite{bhardwaj2017learning,yonetani2021path,chen2020retro}.  Several previous papers focus on \textit{online} graph problems~\cite{sec, LavastidaM0X21, xu2021learning,azar2022online}. The scope of our paper differs from theirs, as we study only offline   problems.

\paragraph{Classic graph algorithms.} There is a vast body of literature addressing graph optimization problems considered in this paper. We only mention a few that are most relevant to this paper. Similar to \cite{dinitz2021faster}, our matching algorithm builds upon the classic Hungarian method. There are other theoretically faster (exact) algorithms for (bipartite) minimum-cost
perfect matching, including \cite{orlin1992new, goldberg1997global, duan2012scaling}. However, these procedures are fairly involved and hard to incorporate predictions. Our learning-based algorithm for single-source negative-length shortest paths  is inspired by \cite{goldberg1995scaling}. In the classic setting, a web of reductions among graph problems were introduced by \cite{gabow1983efficient, gabow1985scaling, gabow1989faster}.  

The learning-based algorithm paradigm has been applied to a number of other problems. See \cref{sec:more} for more related works.

 \subsection{Organization}
 The remainder of the paper proceeds as follows.  We set up some preliminary background and notations in \cref{sec:prelim}. In \cref{sec:faster_matching}, we give an improved algorithm for learning-augmented minimum-cost bipartite matching. We then extend the approach to shortest path in \cref{sec:RE_dual}.  We address the question of learnability in \cref{sec:learnability}. Finally, we provide numerical evaluations of our shortest-path algorithm via reductions in \cref{sec:exp}.

%% file: prelim.tex
\section{Preliminaries}\label{sec:prelim}

\paragraph{Notation.} Let $G = (V, E)$ be a graph of $m$ edges and $n$ vertices. We will specify its directedness in different settings. For a vector $x \in \mathbb{R}^m$, we let $\|x\|_p$ to denote its $p$th norm, for any $p \geq 0$. 

\paragraph{Minimum-Weight Bipartite Perfect Matching.} Let $G = (V, E)$ be a bipartite graph  with non-negative edge costs, and   $C$ be the maximum cost.  The objective of this problem is to find a perfect matching
$M$ with  minimum total cost in $G$.
In the minimum-weight $b$-matching problem, we are also given a demand vector $b \in \mathbb{Z}_+^V$. The goal is to match each vertex $u$ $b_u$ times, with minimum cost.

\paragraph{Maximum Flow.} Given a directed graph $G= (V,E)$ with capacity vector $c \in \mathbb{R}_+^{E}$,  let $s$ and $t$ be distinct vertices of $H$. A feasible $s$-$t$ flow is a vector $f\in\mathbb{R}_+^E$, with each entry representing flow along an edge, such that sum of incoming flow along edges $(v, u)$ equals   sum of outgoing flow along edges $(u,v)$ for all $u \in V \setminus \{s,t\}$ and $f_{uv} \leq c_{uv}$ for all edge $uv$. An $s$-$t$-flow $f$ is maximum if it maximizes the outgoing flow from $s$. For a feasible flow $f$,  $G_f$ denotes its residual graph. A classic procedure for finding maximum flow is Ford-Fulkerson; see \cite{cormen2009introduction}.

%% file: matching.tex
\section{Improved Learning-Based Minimum-Weight   Matching}\label{sec:faster_matching}
The results from \cite{dinitz2021faster} on matching contain two main results: (1) that given predicted duals for a minimum-weight matching problem, there is an efficient near-optimal algorithm to round the duals to feasibility and (2) that after rounding, these feasible predicted duals can be used to quickly find a solution to the primal.
We provide a new approach to the second problem of using a feasible prediction to quickly solve minimum-weight matching that significantly improves upon prior work.

First, we will restate the theorem from \cite{dinitz2021faster} which established an algorithm from using a feasible predicted dual to quickly solve minimum-weight matching.
\begin{theorem}[Theorem $13$ in \cite{dinitz2021faster}]\label{thm:previous-matching}
There exists an algorithm which takes as input a feasible integer dual solution $\hat{y}'$ and finds a minimum-weight bipartite perfect matching in $\widetilde{O}(\min \{m \sqrt{n} \|y^* - \hat{y}'\|_1, mn\})$ time, where $y^*$ is an optimal dual solution.
\end{theorem}

With small modifications to the algorithm and an improved analysis, we present the following improved time complexity.
\begin{theorem}[Faster Matching from Predicted Duals]\label{thm:faster-matching}
There exists an algorithm which takes as input a feasible integer dual solution $\hat{y}'$ and finds a minimum-weight bipartite perfect matching in $O(m\sqrt{n} + (m + n\log n) \|y^* - \hat{y}'\|_0)$ time, where $y^*$ is an optimal dual solution.
\end{theorem}

If the feasible dual is within $O(\sqrt{n})$ of an optimal dual in $\ell_1$ distance (which is the case in which the algorithm from~\cite{dinitz2021faster} attains an improved runtime over the classical algorithm), our algorithm improves upon the time complexity by a factor of
\[
    \sqrt{n} \left(\frac{\|y^* - \hat{y}'\|_1}{\|y^* - \hat{y}'\|_0}\right).
\]
Note that $\|y^*(c) - \hat{y}'(c)\|_0 < \|y^*(c) - \hat{y}'(c)\|_1$ as we are considering only integral duals.

While the algorithm from \cite{dinitz2021faster} improves upon the classic Hungarian algorithm only when $\|y^*(c) - \hat{y}'(c)\|_1 = o(\sqrt{n})$, our algorithm improves upon the Hungarian algorithm as long as $\|y^*(c) - \hat{y}'(c)\|_0 = o(n)$, a much milder condition on the predictions.

As a corollary, when combined with the linear-time rounding procedure from~\cite{dinitz2021faster}, this algorithm gives a fast framework for taking a predicted (possibly infeasible) dual and using it to speed up minimum-weight matching.

\begin{corollary}
There exists an algorithm which takes as input a (possibly infeasible) integral dual solution $\hat{y}$, produces a feasible dual $\hat{y}'$ s.t.\ $\|\hat{y}' - y^*\|_1 \leq 3 \|\hat{y} - y^*\|_1$, and finds a minimum-weight bipartite perfect matching in $O(m\sqrt{n} + (m + n\log n) \|y^* - \hat{y}'\|_0)$ time, where $y^*$ is an optimal dual solution.
\end{corollary}

Our algorithm is given in \cref{alg:fastmatching}. The main difference in the algorithm/analysis to prior work is that they essentially consider running a $O(m\sqrt{n})$ matching algorithm at each step and then reason that the dual variables increase by at least one on each call to the algorithm, getting the $\ell_1$ dependence on the error.  

Our improvements are based on the following observation. If the predicted duals are accurate enough to get improvements over the normal Hungarian algorithm, then the first call to a maximum cardinality matching algorithm will match many edges. Then, we can account for the amount of work we have to do in subsequent iterations by the small number of edges remaining to be matched by via a flow interpretation of the matching problem.
% First, let us restate the algorithm from \cite{dinitz2021faster} (called Algorithm 2 in that paper).

% \begin{algorithm}[H]
% \caption{\label{alg:dinitz} Primal-Dual Scheme for MWPM from \cite{dinitz2021faster}}
% \begin{algorithmic}[1]
% \Procedure{MWPM-PD}{$G = (L \cup R,E),c,y$}
% \State $E' \gets \{ e \in E \mid y_i + y_j = c_{ij}$ \}
% \Comment{Set of tight edges in the dual}
% \State $G' \gets (V, E')$ \Comment{$G'$  containing only tight edges}
% \State $M \gets$ \text{ Maximum cardinality matching in }$G'$
% \While{$M$ is not a perfect matching}
% \State Find $S \subseteq L$ such that $|S| > |\Gamma(S)|$ in $G'$
% \Comment{Exists by Hall's Theorem}
% \Statex \Comment{Can be found in $O(m+n)$ time}
% \State $\eps \gets \min_{i \in S,j \in R\setminus \Gamma(S)} \{ c_{ij} - y_i - y_j \}$
% \State $\forall i \in S$, $y_i \gets y_i + \eps$
% \State $\forall j \in \Gamma(S)$, $y_j \gets y_j - \eps$
% \State Update $E',G'$
% \State $M \gets$ \text{ Maximum cardinality matching in }$G'$
% \EndWhile
% \State Return $M$
% \EndProcedure
% \end{algorithmic}
% \end{algorithm}

% The algorithm we consider does the same first 3 steps and then considers the min-cost max-flow view of the min-cost perfect matching problem as in the Hungarian algorithm described in \cite{ahuja1993networkflows}. 

\begin{algorithm}[ht]
\caption{\label{alg:fastmatching} Faster Primal-Dual Scheme for MWPM}
\begin{algorithmic}[1]
\Procedure{MWPM-PD$++$}{$G = (L \cup R,E),c,y$}
\State $E' \gets \{ e \in E \mid y_i + y_j = c_{ij}$ \}
%\Comment{Set of tight edges in the dual}
\State $G' \gets (V, E')$ %\Comment{$G$  containing only tight edges}
\State $M \gets$ \text{ Maximum cardinality matching in }$G'$
\State Give all edges in $E$ unit capacity and direct them from left to right \Comment{Flow representation}
\State Add nodes $s, t$ to $G$ along with unit capacity, zero cost edges $(s, i)$ for all $i \in L$ and $(j, t)$ for all $j \in R$
\State Associate a flow $f$ with $M$ s.t. $\forall (i,j) \in M$, $f_{si} = f_{ij} = f_{jt} = 1$ and otherwise $f_e = 0$
\State $z_i \gets -y_i \quad \forall i \in L$
\State $z_j \gets y_j \quad \forall j \in R$
\State $c'_e \gets c_e + z_i - z_j \quad \forall e=(i,j) \in E \text{ s.t. } i,j \notin \{s,t\}$ and $c'_e \gets 0$ for all other edges %\Comment{Reduced costs}
\While{$f$ has flow value less than $n$}
\State $z_u \gets z_u + d(s,u) \; \forall u \in L \cup R$ where $d(\cdot, \cdot)$ is shortest path distance in $G_f$ w.r.t.\ $c'$
\Comment{Dijkstra}
\State $c'_e \gets c_e + z_i - z_j \quad \forall e=(i,j) \in E \text{ s.t. } i,j \notin \{s,t\}$ 
\State $E_f' \gets \{e \in E_f | c'_e = 0\}$.
\State $G_f' \gets (V, E_f')$
\State $g \gets$ Maximum flow in $G_f'$
\Comment{Ford-Fulkerson}
\State Augment along $g$ in $G_f$
\EndWhile
\State Return $\{e=(i,j) \in f: i \in L, j \in R, f_e = 1\}$
\EndProcedure
\end{algorithmic}
\end{algorithm}

The formal analysis of the algorithm is somewhat technical and appears in \cref{sec:o3}, where we prove \cref{thm:faster-matching}.

% Finally, we remark that our result can be extended to $b$-matching. Using a similar appraoch to the analysis, we improve upon~\cite{dinitz2021faster} by more than an $O(n)$ factor for a large parameter regime. See \cref{sec:bmatch} for details.

\paragraph{Extension to $b$-matching}
We extend the improvements for learning-based minimum-weight perfect bipartite matching to the more general problem of minimum-weight perfect $b$-matching.
For two sets of dual variables over the vertices $y$ and $z$, we will use as a distance measure the weighted $\ell_p$ error:
\[
\|y - z\|_{b, p} = \sum_{i \in V} b_i |y_i - z_i|_p.
\]
We will restate a theorem from \cite{dinitz2021faster}.
\begin{theorem}[Theorem $31$ in \cite{dinitz2021faster}]\label{thm:previous-b-matching}
There exists an algorithm which takes as input a feasible integer dual solution $\hat{y}'$ and finds a minimum-weight perfect b-matching in $O(mn \|y^* - \hat{y}'\|_{b, 1})$ time, where $y^*$ is an optimal dual solution.
\end{theorem}

Using the same algorithm (\cref{alg:mwbm-pd} shown in \cref{sec:bmatch}), but with an improved analysis, we show the following improved runtime.
\begin{theorem} \label{thm:improved-b-matching}
There exists an algorithm which takes as input a feasible integer dual solution $\hat{y}'$ and finds a minimum-weight perfect b-matching in $O(mn +  m\|y^* - \hat{y}'\|_{b, 0})$ time, where $y^*$ is an optimal dual solution.
\end{theorem}
As before, since the duals are integral, $\|y^* - \hat{y}'\|_{b, 0} \leq \|y^* - \hat{y}'\|_{b, 1}$. Note that this runtime improves upon prior work by a factor of
\[
\min\left\{n \frac{\|y^* - \hat{y}'\|_{b, 1}}{\|y^* - \hat{y}'\|_{b, 0}}, \|y^* - \hat{y}'\|_{b, 1}\right\}.
\]
The full details and proof are in \cref{sec:bmatch}.

%% file: shorestpath_duals.tex
\section{Fast Learning-Based Shortest Paths}\label{sec:RE_dual}

In this section, we introduce the reduced edge length duals and how to round them efficiently given predictions. Reduced edge length duals are defined as follows.

\begin{definition}[Reduced Edge Length Duals (RE Duals)]\label{def:reduced_edge}
Let $G = (V,E)$ with $|V| = n, |E| = m$, denote a directed graph and $\ell: E \rightarrow \mathbb{Z}$ denote the length of the edges, which may be negative. $y \in \mathbb{Z}^V$ is a valid or feasible reduced edge length dual (RE Dual) if 
\[\ell_y(u,v) := \ell(u,v) + y_u - y_v \ge 0 \]
for all edges $e = (u,v) \in E$.
\end{definition}

It is natural to study these duals as they appear in many fundamental combinatorial optimization problems. For example, consider the dual linear program for the shortest paths problem on the graph $G$ (where we wish to compute the shortest path from vertex $s$ to $t$). It is given by:

\begin{align*}\label{eqn:shortest_path_dual}
    \max & \quad  y_t \\
    \text{s.t.} &\quad  y_v - y_u \leq  \ell(u,v)\\
    &\quad  y_s = 0.
\end{align*}

Note that the constraints $y_v - y_u \leq  \ell(u,v)$ exactly encode $\ell_y(u,v) \ge 0$ in  \cref{def:reduced_edge}. Furthermore, given a valid dual solution $y$ to the dual linear program, one can quickly compute the shortest paths in near linear time via an application of Dijkstra's algorithm since all reduced edge lengths are non negative by   \cref{def:reduced_edge}. This is because the sum of the lengths of edges along any path $(v_1, v_2, \cdots, v_k)$ is the same up to an additive term $y_{v_1} - y_{v_k}$ due to telescoping. Thus, this transformation preserves the identity of shortest paths from a starting vertex. Furthermore, many shortest paths algorithms on general graphs, such as the Bellman-Ford algorithm, also implicitly calculate the dual $y$: in the Bellman-Ford algorithm, the dual can be constructed in linear time after the algorithm terminates.

Now suppose predictions $\hat{y}: V \rightarrow \mathbb{Z}$ for the duals $y$ are given. The main result of this section is that there exists an efficient algorithm, \cref{alg:round_shortest_paths}, which outputs a feasible $\hat{y}'$ according to \autoref{def:reduced_edge}. 

\begin{theorem}[Fast Shortest-Path from Predicted Duals]\label{thm:RE_dual_rounding}
Let $\hat{y}: V \rightarrow \mathbb{Z}$  be predicted duals and let $y^*: V \rightarrow \mathbb{Z}$ be a feasible set of reduced edge length duals according to   \cref{def:reduced_edge} such that $\|\hat{y}-y^*\|_1$ is minimized. \cref{alg:round_shortest_paths} returns a feasible $\hat{y}': V \rightarrow \mathbb{Z}$ in time \[O(m \min\{\|\hat{y}-y^*\|_1 \cdot \|\hat{y}-y^*\|_{\infty}, \sqrt{n} \log(\|\hat{y}-y^*\|_{\infty})\}).\]
\end{theorem}

 If we define reduced edge lengths according to the predicted dual $\hat{y}$, it is likely that the non-negativity constraint of some edges become violated, i.e., $\ell_{\hat{y}}(e) < 0$. The goal of \cref{alg:round_shortest_paths} is to modify some coordinates of $\hat{y}$ to fix these negative edge weights. The algorithm uses a key subroutine of Goldberg's algorithm on shortest paths \cite{goldberg1995scaling}. It proceeds by mending negative edges  by reducing the dual value of one of their endpoints. At every iteration, we greedily maximize the number of dual variable which are updated. The vertices which are updated are picked through a layering structure utilized in \cite{goldberg1995scaling}. \cref{alg:round_shortest_paths} presents the formal details. 

Note that we implicitly assume the given graph $G$ with edge lengths given by $\ell$ does not have a negative weight cycle. This is a necessary assumption since otherwise, there exists no valid RE Duals for $G$: the length of a cycle under any valid dual $y$ must be non-negative by definition but the cost of any cycle is the same under $\ell$ and $\ell_y$ due to telescoping which leads to a contradiction if $\ell$ induces a negative weight cycle.

\begin{algorithm}[ht]
\caption{\label{alg:round_shortest_paths} Rounding Predictions for Reduced Edge Length Duals}
\begin{algorithmic}[1]
\State \textbf{Input:} Graph $G = (V,E)$, predicted duals $\hat{y}: V \rightarrow \mathbb{Z}$
\Procedure{Round-RE-Duals}{$G, \hat{y}$}
\While{there exists an edge $e$ such that $\ell_{\hat{y}}(e) < 0$}
\State $G^{-} = (V,E^{-}) \gets$ subgraph of $G$ that have weight at most $0$ under $\ell_{\hat{y}}$
\State Contract all strongly connected components in $G^-$
\Comment{All edges connecting vertices in the same strongly connected component are $0$ \cite{goldberg1995scaling}}
\State Add a vertex $x$ to $G^{-}$ and connect it with zero length edges to all of $V$
\State $L_i \gets \{v \in V | d(x,v) = -i\}$ 
\Comment{$d$ is graph distance in $G^{-}$ using reduced edge lengths given by $\ell_{\hat{y}}$}
\State $i^* \gets \argmax_i |L_i|$
\State Lower the value of $\hat{y}_v$ for all vertices in $\cup_{t \ge i^*} L_t$ by $1$ \label{ln:modified-dual}
\EndWhile
\State Return $\hat{y}$
\EndProcedure
\end{algorithmic}
\end{algorithm}

The analysis of the algorithm and the proof of \cref{thm:RE_dual_rounding} appear in \cref{sec:o4}.
 The theorem also implies an algorithm for all-pair shortest paths; see \cref{sec:apsp} for details.

%% file: reductions.tex
\section{A General Framework for Learning-Based Reductions}

In this section, we introduce a general framework for obtaining learning-augmented algorithms via reductions. Suppose we have an oracle which provides hints or a warm start to instances of problem $P_1$. If we are instead interested in solving instances of another problem $P_2$, we can hope to transform our instance at hand to an instance of $P_1$ in order to utilize the available predictions. If there exists an efficient reduction from $P_2$ to $P_1$, we can indeed use this reduction to transform our instance of $P_2$ to that of $P_1$, use the hints available for $P_1$ to efficiently solve our new problem, and use the solution found to solve our original instance of $P_2$. This will be the basis of our framework for learning-based reductions.

Why is such a framework useful? First, hints might be easier to learn for problem $P_1$ or there may not be a natural notion of hints for instances of $P_2$. In addition, there might already exist a learning-based algorithm for $P_1$ which efficiently utilizes hints. Therefore, using reductions from other problems to $P_1$ would eliminate the need to create new algorithms and thereby increasing the power and usability of the existing learning-based algorithms. 

We formally define reductions as follows.

\begin{definition}[Reductions]Let $P_1$ and $P_2$ be two problem instances. We say that $R: P_2 \rightarrow P_1$ is a reduction from $P_2$ to $P_1$ if for any instance $I \in P_2$, $R(I)$ maps to an instance $I'$ of $P_2$. 
Furthermore, Furthermore, there exists mapping which takes a solution of $I'$ and converts it to a solution for $I$.
%\ali{In this framework I assume we are only considering ``exact'' solutions. If yes, we can mention it.}
%Furthermore, solving $I'$ gives us a solution for $I$.
\end{definition}

Note that the definition of reduction by itself is not quite useful: by the Cook-Levin theorem, any problem in the complexity class P can be reduced to 3SAT. However in this paper, we are interested in \textit{efficient} reductions which take linear or almost linear time in the size of the input. Therefore, such reductions would be extremely fast to execute in practice and the final algorithm of solving instances of $I$ of $P_2$ via solving instances of $P_1$ using a learned oracle would overall be faster than solving $I$ with no hints. 

\subsection{General Framework}
Our framework is given in   \cref{alg:reduction}. Note that there, $\mathcal{A}$ is an existing algorithm which solves instances of problem $P_1$ using hints given by a predictor $y : P_1 \rightarrow \R^d$, i.e., the hints are $d$ dimensional vectors.

\begin{algorithm}[H]
\caption{\label{alg:reduction} General Reduction Framework for Learning-Based Algorithms}
\begin{algorithmic}[1]
\State \textbf{Input:} Problem instance $I \in P_2$. Predictor $y : P_1 \rightarrow \R^d$, reduction $R : P_2 \rightarrow P_1$, Algorithm $\mathcal{A}(P_1, y(P_1))$
\Procedure{Reduction-Solve}{$P_1, P_2, y, R, \mathcal{A}$}
\State $I' \gets R(I)$ \Comment{Use reduction $R$ to get an instance of $P_1$}
\State $\hat{y} \gets y(I')$ \Comment{Get hints for instance $I'$ using predictor $y$}
\State Execute $\mathcal{A}(I', \hat{y})$ \Comment{Solve instance $I'$ using hints and learning-based algorithm $\mathcal{A}$}. 
\State Return solution to $I$ using solution for $I'$ given by  $\mathcal{A}(I', \hat{y})$.
\EndProcedure
\end{algorithmic}
\end{algorithm}

Note that Step $6$ in \cref{alg:reduction} would depend on the instances $P_1, P_2 $ and the reduction $R$. In some cases, some post-processing the solution $\mathcal{A}(I', \hat{y})$ could be required. In the examples we study in this paper, both this step and the reduction $R$ are efficient. We give concrete instantiations of \cref{alg:reduction} in~\cref{sec:reductions}.

Note that we still need to understand the learnability of the hints $\hat{y}$ in Step $4$ of \cref{alg:reduction}: even if there exists an efficient algorithm $\mathcal{A}$ for problem $P_1$, we might not have a predictor $y$ at hand. Note that in \cite{dinitz2021faster}, the question of learnability of predictors was tackled by assuming access to multiple instances of a particular problem class drawn from some distribution. In our case, we might have lots of training data on instances of problem $P_2$ but our goal is to train a predictor $y$ for $P_1$ in hopes of utilizing $\mathcal{A}$. To do so, we can just go through the reduction $R$ to get samples of problem instances drawn from $P_1$. Note that the distribution on these problems will be different than that on $P_2$. We introduce general learnability results which imply one can learn a good predictor $y$. For details, see ~\cref{sec:learnability}. 

\subsection{Reductions and Their Implications}\label{sec:reductions}

We now present one application of the reduction framework outlined in the previous section. 
We demonstrate three more reductions, for degree constrained subgraph, minimum-cost $0$-$1$ flow, and graph diameter, in \cref{sec:more-red}.

\paragraph{Shortest Path from Matching.}
We leverage the following reduction from shortest path (with negative edge lengths) to  maximum-weight perfect matching on bipartite graphs, due to \cite{gabow1985scaling}. Given a directed graph $G$ with edge lengths given by $\ell$, construct a weighted bipartite graph $H = (L, R, E)$. 
\begin{itemize}
    \item For each vertex $u \in G$,   make two copies $u_1 \in L$ and $u_2 \in R$. 
    \item  For each arc $(u,v) \in G$, create an edge $e=(u_1,v_2)$ of weight $-\ell(e)$ in $H$. 
    \item Finally, create an edge $(u_1,u_2)$ of weight $0$ for each vertex $u \in G$. 
\end{itemize}
Now suppose we find the maximum weight perfect matching of $H$ and its corresponding dual variables $y_{u_i}$ for all $u \in G$ and $i\in \{1,2\}$.  By the construction, we immediately have:
\begin{lemma}\label{lem:dual_transformation}
The maximum weight perfect matching of $H$ has positive weight if and only if 
the graph $G$ has negative cycles.
\end{lemma}
Otherwise, we can let $\pi_u = y_{u_1}$ for each $u\in G$. Then by feasibility of $y$, we have $$-\ell(e)\leq y_{u_1} + y_{v_2} \leq \pi_u -\pi_v, $$ for any $e= (u,v) \in G$. It follows that $\pi$ is a feasible dual for the shortest path problem on $G$, i.e., it satisfies   \cref{def:reduced_edge}.

Observe that the graph $H$ contains $m+n$ edges. By using the run-times for faster matching from \cite{dinitz2021faster} as well as our runtime of Section \cref{sec:faster_matching}, we have the following corollary due to \cref{alg:reduction}.

\begin{theorem}\label{thm:shortestpath_reduction}
Given a shortest path problem on input graph $G=(V,E)$ with $n$ vertices and $m$ edges, there exists an algorithm which takes takes as input a predicted dual solution $\hat{y}$ to an instance of maximum weight perfect matching derived from $G$, near-optimally rounds the dual to a feasible solution $\hat{y}'$, and finds feasible reduced edge length duals for $G$ in time $O(m\sqrt{n} + (m + n\log n) ||y^* - \hat{y}'||_0)$.
\end{theorem}

\begin{remark}
Recall that in \cref{sec:RE_dual} we derived an alternative runtime of $$O( m \min\{\|\hat{y}-y^*\|_1 \cdot \|\hat{y}-y^*\|_{\infty}, \sqrt{n} \log(\|\hat{y}-y^*\|_{\infty}))\}$$ for the problem of rounding a predicted dual $\hat{y}$ to a feasible dual $\hat{y}'$ to satisfy the reduced edge property of Definition \ref{def:reduced_edge}. These results are incomparable since \cref{thm:shortestpath_reduction} uses dual predictions for matchings on a transformed graph whereas \cref{thm:RE_dual_rounding} uses predictions for shortest path duals (RE duals) on the original graph.
\end{remark}

%% file: learnable.tex
\section{General Learnability of Hints}\label{sec:learnability}
We now present two general learnability theorems on vector-valued hints for graph optimization problems.  We consider the model where  the edge weights (or capacities) are drawn  from a distribution, while the vertex set is the same. The goal is to learn a hint vector in $\mathbb{R}^d$ that is close to the \textit{optimal hint} on average (in $\ell_1$ or $\ell_\infty$ norm), given i.i.d.\ samples of the edge weights. We  require  no assumption either on the edge weights distribution or on the notion of optimality of a hint.  Indeed, the latter needs to depend on particular problems. For a variety of graph problems, though, the optimal hint could be taken as the optimal dual solution of certain LP relaxation. 

Throughout the section, we assume that the edge weights $c$ are drawn from an unknown distribution of $\mathcal{D}$.  We search for a hint within a range $\mathcal{H} \subseteq \mathbb{R}^d$.

\subsection{Learnability from Bounded Pseudo-Dimension}\label{sec:learnability_pd}

%Let $C = \max_{c\sim \mathcal{D}} \max_e |c_e|$ be the upper bound on the maximum value of any edge weight. 
For a fixed problem,  given edge weight $c\in \mathbb{R}^m$, and graph $G$, let $h^*(c)$ denote an optimal hint with respect to the instance $(G,c)$. We consider $\ell_1$ and $\ell_\infty$ loss:
\begin{align}
        \ell_1(h,c) = \| h^*(c)  - h \|_1 = \sum_{i=1}^d \left|h^*_i(c) - h_i\right|,\\
        \ell_\infty(h,c) = \| h^*(c)  - h \|_\infty = \max_i \left|h^*_i(c) - h_i\right|.
\end{align}
The goal of the algorithm is to find a hint $\widehat{h} \in \mathcal{H}$ such that the expected loss $\E_{c\sim \mathcal{D}} \ell(h,c)$ is minimized, for $\ell = \ell_1$ or $\ell_\infty$. Let $h^* \in \argmin_{h\in \mathcal{H}} \E_{c\sim \mathcal{D}} \ell(h,c)$. %Finally, we remark that the general learning algorithms we propose here do not concern with   feasibility of the hints, as  the constraints are determined by   particular instances.  Rather, the objective is simply to achieve small population loss.
\paragraph{$\bm{L}_1$ Loss.}
Our first result is a straightforward abstraction of the main learnability theorem of~\cite{dinitz2021faster},  for the $\ell_1$ loss.
In particular, we show that one can find a hint vector $\widehat{h} \in \mathbb{R}^d$ that approximately minimizes the population loss $\E_{c\sim \mathcal{D}} \ell_1(h,c)$, under the following conditions:

\begin{theorem}[$\ell_1$-learnability; see also Theorem 14 of \cite{dinitz2021faster}]\label{thm:gen-learn-uni}
For any graph problem with optimal hint $h^*(c)\in\mathcal{H}$ for $c\sim \mathcal{D}$, assume that 
\begin{itemize}
    \item (bounded range) for any $h \in \mathcal{H}$ we have $h_i\in [-M, M]$  for all $i$, for some $M$; and
    \item (efficient optimization)  there exists a polynomial time algorithm that finds
 a hint vector $h \in\mathcal{H}$ that minimizes $\sum_{i=1}^s \|h^*(c_i) - h\|_1$, given i.i.d.\ samples $c_1,c_2,...,c_s \sim \mathcal{D}$.
\end{itemize}
Then there is a polynomial-time algorithm that given $ s=O\left(\left(\frac{dM}{\epsilon}\right)^{2}(d \log d+\log (1 / \delta))\right)$ samples returns a hint $h \in \mathcal{H}$ such that $\E_{c\sim \mathcal{D}} \ell_1(h,c) \leq \E_{c\sim \mathcal{D}} \ell_1(h^*(c),c) +\epsilon $ with probability at least $1- \delta$. 
 \end{theorem}
\paragraph{$\bm{L}_\infty$ Loss.} 
We now give a learnability result for the $\ell_\infty$ loss.
\begin{theorem}[$\ell_\infty$-learnability]\label{thm:gen-learn-uni-infty}
For any graph problem with optimal hint $h^*(c)\in\mathcal{H}$ for $c\sim \mathcal{D}$, assume that 
\begin{itemize}
    \item (bounded range) for any $h \in \mathcal{H}$ we have $h_i\in [-M, M]$  for all $i$, for some $M$; and
    \item (efficient optimization)  there exists a polynomial time algorithm that finds
 a hint vector $h \in\mathcal{H}$ that minimizes $\sum_{i=1}^s \|h^*(c_i) - h\|_\infty$, given i.i.d.\ samples $c_1,c_2,...,c_s \sim \mathcal{D}$.
\end{itemize}
Then there is a polynomial-time algorithm that given $ s=O\left(\left(\frac{dM}{\epsilon}\right)^{2}(d+\log (1 / \delta))\right)$ samples returns a hint $h \in \mathcal{H}$ such that $\E_{c\sim \mathcal{D}} \ell_\infty(h,c) \leq \E_{c\sim \mathcal{D}} \ell_\infty(h^*(c),c) +\epsilon $ with probability at least $1- \delta$. 
 \end{theorem}
 The proofs of the two theorems appear in \cref{sec:o6}.

\subsection{Learnability from Arithmetic Complexity}
We   give  an alternative argument for learning predictions. Informally, we show that good predictions can be learned efficiently   if the loss function can be `computed efficiently'. This provides a more general framework that goes beyond $\ell_1$ or $\ell_\infty$ norm error. See \cref{sec:arith} for formal details.

%% file: experiment.tex
\section{Empirical Simulations}\label{sec:exp}
We demonstrate the applicability of our learning-based reductions framework with a real world case study on foreign exchange markets. The reduction we focus on is the 
general shortest paths to bipartite matching reduction outlined in Section \ref{sec:exp}. We focus our evaluation on this task since prior work in \cite{dinitz2021faster} has already demonstrated the empirical advantage of learning-based methods for bipartite matching.

Our graph dataset is constructed as follows. We have a weighted directed graph where nodes represent countries and all possible directed edges between all pairs are present. The weight of the directed edge from country $i$ and $j$ represents the average monthly exchange rate between the currencies of the two countries, i.e., the amount of currency $j$ we can obtain starting from one unit of currency $i$, as set by the foreign exchange rate market\footnote{Dataset scraped from \url{https://fxtop.com/en/historical-exchange-rates.php}}. We transform these weights by taking the natural logarithm and negating the weight. This implies that the shortest path from country $i$ to country $j$ on the transformed graph represents the \emph{optimal} way to convert one unit of currency $i$ to the currency of $j$, i.e., the set of conversions which maximize the amount of currency $j$.

\paragraph{Experimental Setup.} We first describe our training dataset. We construct the graph described above for each month of the year $2019$ where we use the average monthly exchange rates as edge weights before performing the transformation. Our testing dataset are similarly constructed graphs for each month of $2020$ and $2021$. For each graph, we construct the reduction from shortest paths to matching outlined in  \cref{sec:reductions}. By   \cref{lem:dual_transformation}, the output of the maximum weight perfect matching on the bipartite graph obtained via the reduction  gives us feasible reduced edge length duals which we can be subsequently use to solve shortest paths in nearly linear time. The resulting bipartite graphs have 
$\sim 500$ vertices each and $\sim 5 \cdot 10^4$ edges.

We use the code from \cite{dinitz2021faster} \footnote{Available  at \url{ https://github.com/tlavastida/LearnedDuals}} for the maximum weight bipartite matching algorithm. As in \cite{dinitz2021faster}, we measure the runtime in terms of the number of steps used by the Hungarian algorithm to solve the matching instances derived from our training and test graph datasets when we initialize the algorithm with predicted duals versus when we start the algorithm ``from scratch.'' %Measuring the number of Hungarian steps is desirable since it is a software and architecture-free metric, unaffected by the use of specialized libraries, hardware, or computing infrastructure or frameworks, which vary for graph algorithms. Furthermore, such as measure is compatible with prior literature (e.g., \cite{dinitz2021faster}).

We instantiate predictions in two distinct ways, similar to the methodology of \cite{dinitz2021faster}: (a) first, we consider the \emph{batch} version where we compute the optimal dual variables in the training set, take their median, and use these as the predicted dual variables for each of the graphs in the test set. (b) The second method is the \emph{online} version where we use the optimal dual variables from the graph for the prior month in the test set as initialization for the current month. 

% Both methods have advantages: the batch method is less sensitive to outliers in data and small scale distribution shifts. In addition, our theoretical bounds on PAC learnability derived in   \cref{sec:learnability} apply to the batch setting.
% By contrast, the online method makes use of domain knowledge that foreign exchange rates generally change slowly over time, making last months data the most pertinent for this month.
%The online method is better if the distribution of inputs stable and input instances which are close in time are similar. This allows for a more precise calculation of dual predictions.

\paragraph{Results.} Our results are shown in Figures \ref{fig:alg} and \ref{fig:dual_excess}. Figure \ref{fig:alg} shows up to an \emph{order of magnitude} reduction in the number of iterations taken by the learning-augmented algorithm versus the classical and widely used Hungarian algorithm. As expected, the online method performs slightly better than the batch version as it is able to offer more accurate predictions for the next graph instance. This is very intuitive: it is rare for the foreign exchange market to experience drastic shifts over the span of one month since such a shift implies a major global event.

Our results also validate the dependence of prediction error derived in our theoretical bounds. In Figure \ref{fig:dual_excess}, we plot the excess dual objective, defined as $\sum_e y^*_e - \sum_e \hat{y}_e$ where recall that $y^*$ represents the optimal dual variables and $\hat{y}$ denotes the predictions, versus the number of steps saved in the Hungarian algorithm in our batch setting; we obtained a qualitatively similar result for the online setting. We see there is a direct linear relationship between the excess dual objective, which represents the prediction error, and the decrease in runtime, measured by the number of Hungarian iterations saved. Note that we removed three outlier points from Figure \ref{fig:dual_excess} which represent data from October 2021 to December 2021. The outlier points showed a large excess dual as well as large savings in runtime (which can be inferred from Figure \ref{fig:alg}). We hypothesize that this is because of a distribution shift which occurred during these months in the foreign exchange markets. Indeed, examining the conversion rate from the Euro to US Dollars for example, we see a $3\%$ decrease in the exchange rate which represents the biggest decrease in the time frame of our training and test dataset. This can  be explained by global events such as the rise of the Omicron strain or concerns about increased inflation.

In addition to extending and complementing the experimental results of \cite{dinitz2021faster}, we summarize our results in the following points: 
(a) Our theory is predictive of experimental performance. Both figures demonstrates that better predictions imply better empirical runtime. In addition, Figure \ref{fig:dual_excess} demonstrates a direct relationship between prediction error and runtime, as implied by our theoretical bounds. (b) The reduction framework is efficient to carry out and execute in practice. (c) Learning augmented graph algorithms can be applied to real world datasets varying over time such as in the analysis of graphs derived from the foreign exchange rates market.

\ifnum\value{num}>0 {

\begin{figure}
\centering
\begin{minipage}{.5\textwidth}
  \centering
  \includegraphics[width=7.5cm]{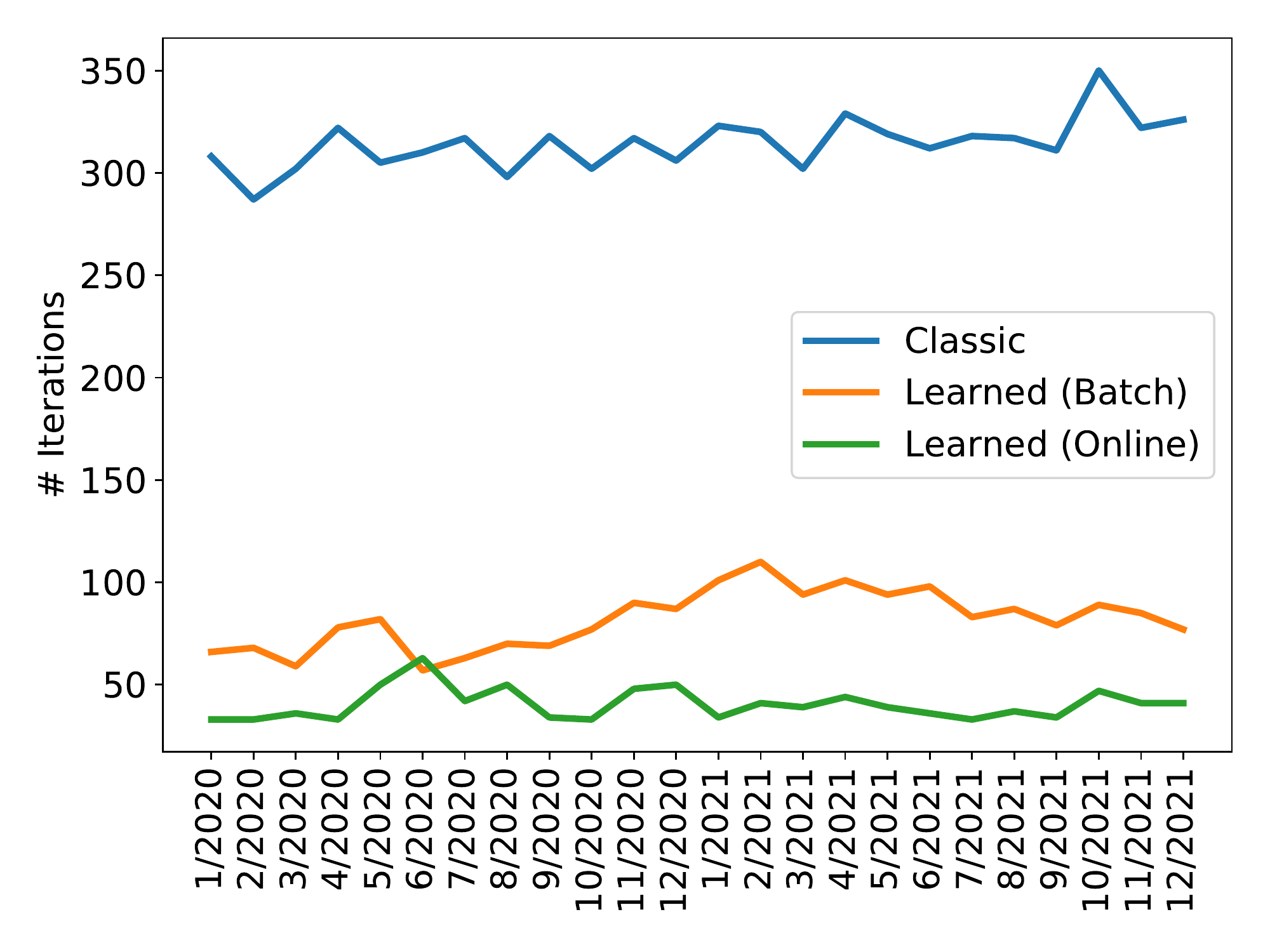}
  \captionof{figure}{Comparison of the classical Hungarian \\ algorithm (blue) versus learning-augmented \\ algorithms. Predictions lead to up to an order \\ magnitude reduction in number of iterations.}
  \label{fig:alg}
\end{minipage}%
\begin{minipage}{.5\textwidth}
  \centering
  \includegraphics[width=7.5cm]{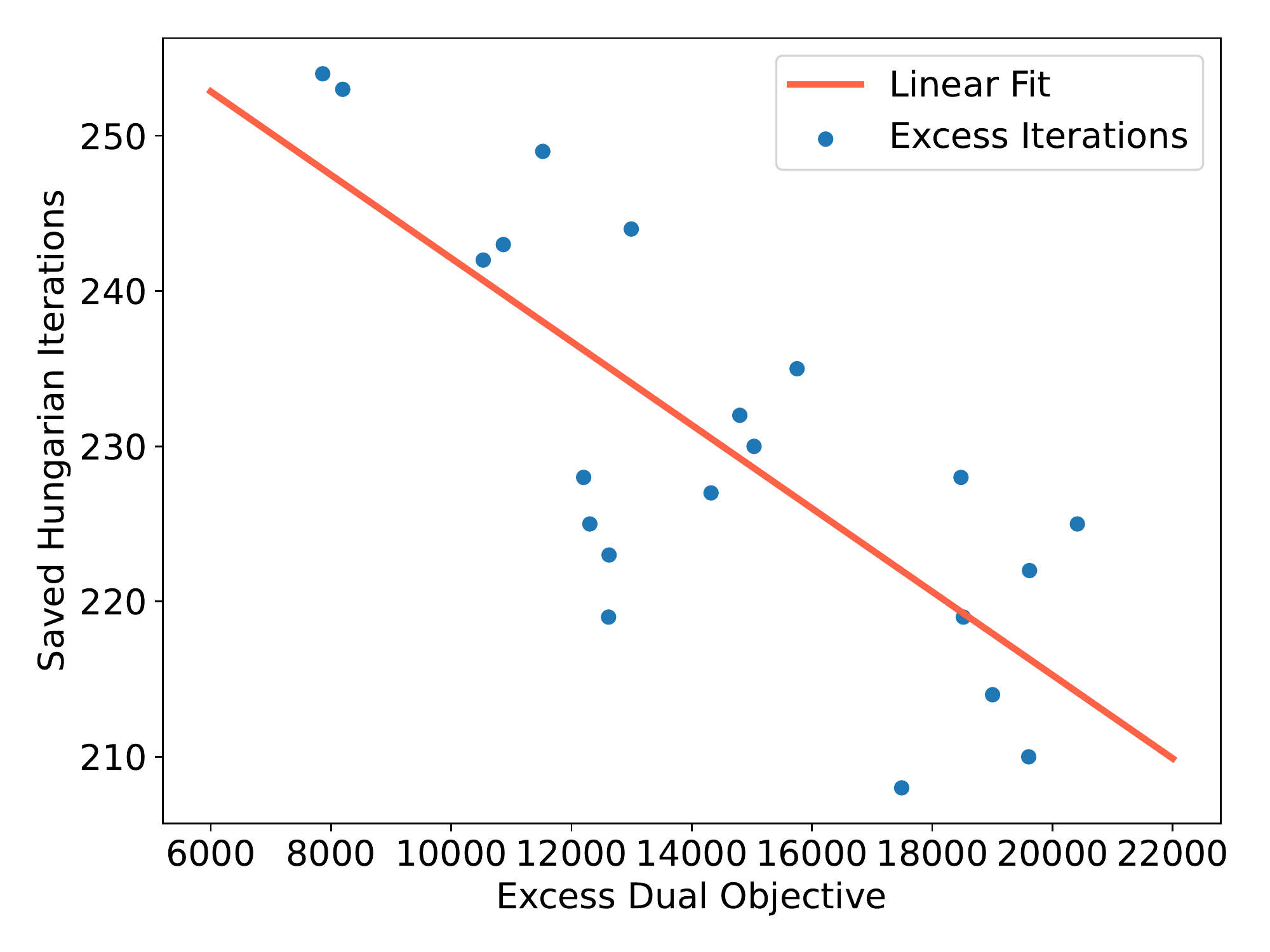}
  \captionof{figure}{Excess dual objective versus the number \\ of saved Hungarian iterations in the batch version. \\ There is a negative correlation between the \\ excess dual and the $\#$ of saved iterations}
  \label{fig:dual_excess}
\end{minipage}
\end{figure}

} \else {

\begin{figure}[h]
\centering
\includegraphics[width=0.8\columnwidth]{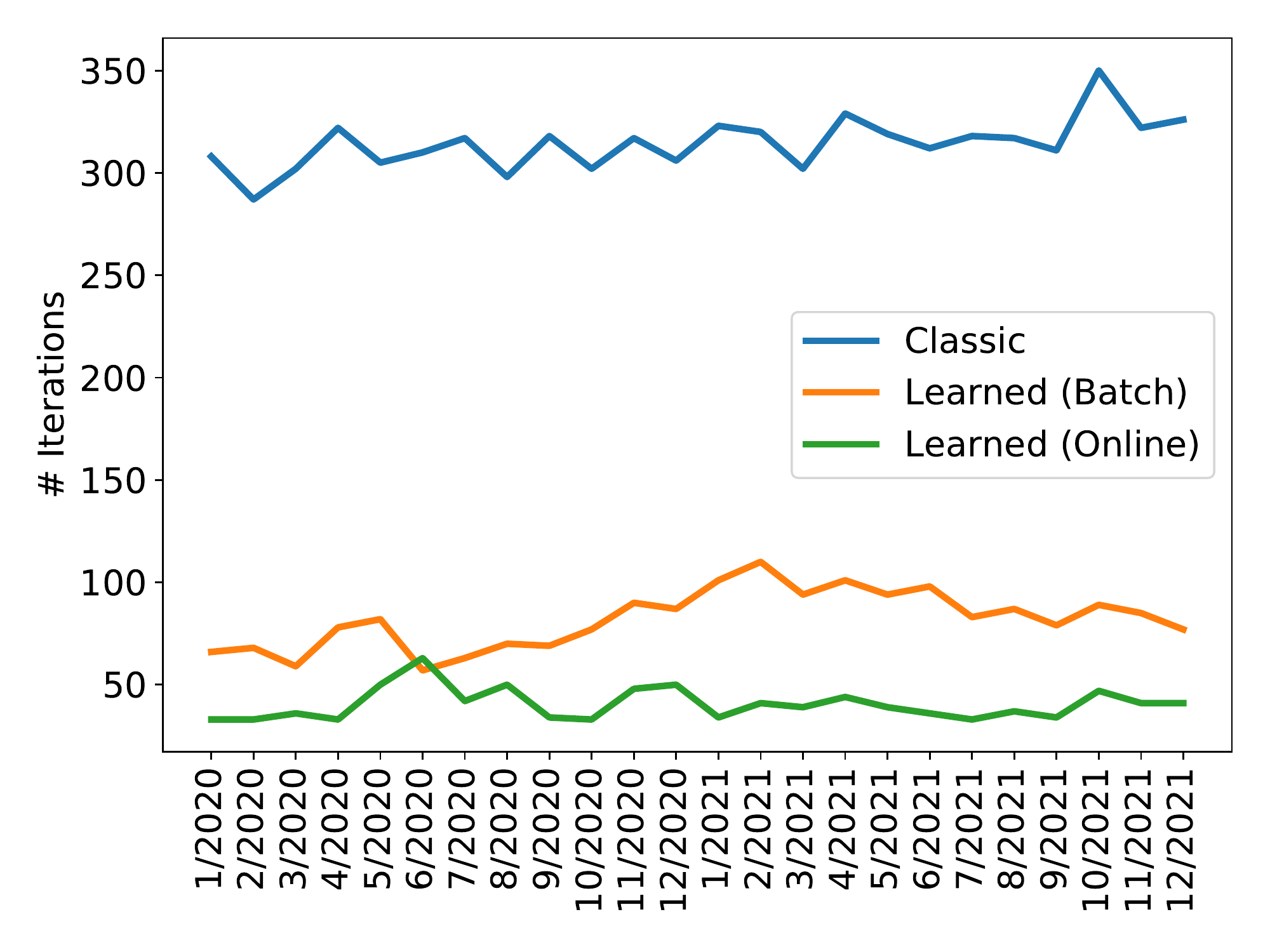}
\vspace{-1mm}
\caption{\label{fig:alg} Comparison of the classical Hungarian algorithm (blue) versus learning-augmented algorithms. Predictions lead to up to an order magnitude reduction in number of iterations.}
\vspace{-2mm}
\end{figure}

\begin{figure}[h]
\vspace{-2mm}
\centering
\includegraphics[width=0.8\columnwidth]{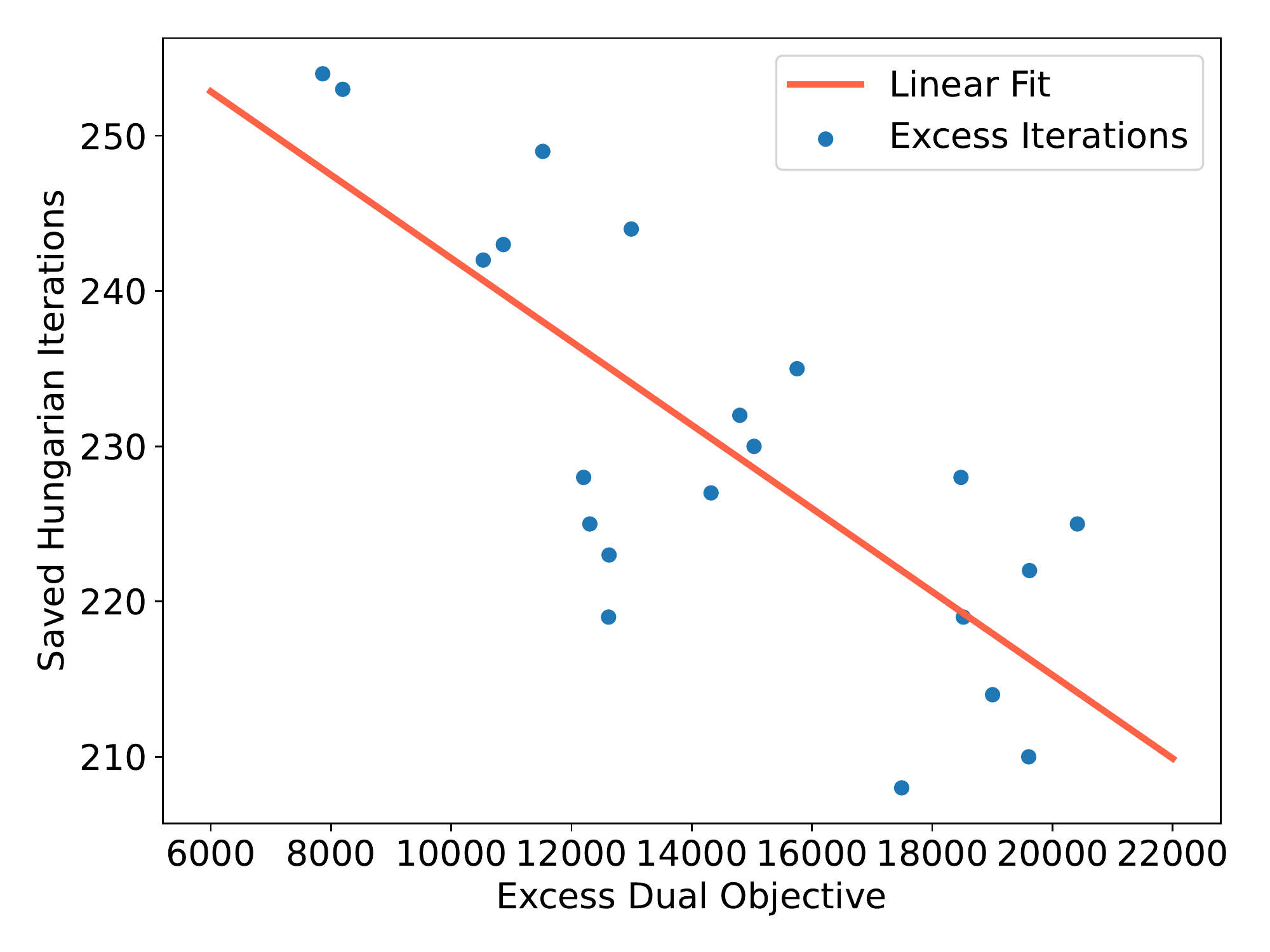}
\vspace{-1mm}
\caption{\label{fig:dual_excess} Excess dual objective versus the number of saved Hungarian iterations in the batch version. There is a negative correlation between the excess dual and the $\#$ of saved iterations.}
\end{figure}

}\fi

%% file: icml22/other.tex
\section{More Related work}\label{sec:more}
Learning-augmented approaches have   found success in a wide array of algorithmic tasks.
This includes improving classic space complexity in streaming algorithms~\cite{hsu2018,indyk2019learning,cohen2020composable,Jiang2020, EINR+2021,du2021putting}, and achieving better competitive ratios in online algorithms~\cite{mahdian2012online, esfandiari2015online,lykouris2018competitive, purohit2018improving, gollapudi2019online,rohatgi2020near,wei2020better, lattanzi2020online,bamas2020primal, chen2021,anand2020customizing,WeiZ20,diakonikolas2021learning, anand2021regression,bansal2022learning}. Other application domains include data structures~\cite{kraska2017case,ferragina2020learned,mitz2018model,rae2019meta,vaidya2021partitioned}, similarity search~\cite{wang2016learning,dong2020learning}, and machine scheduling~\cite{mitzenmacher2020scheduling, ahmadian2022robust, lattanzi2020online, im2021non}. 

A number of works   study deep and reinforcement learning for combinatorial optimization   and integer programming~\cite{bello2016neural,khalil2016learning2,khalil2017learning2,khalil2017learning,pmlr-v80-balcan18a,nazari2018reinforcement,NIPS2018_7335,kool19,selsam2018learning,mao2019learning,amizadeh2018learning,alabi2019learning,zhu2021network,alomrani2021deep}. Most of the work in this direction are empirical in nature. See \cite{mazyavkina2021reinforcement,bengio2021machine}
 for two recent surveys.
 
There are also recent works on a related area of data driven algorithm design whose focus is to learn a generalizable algorithm from a family using a small number of samples rather than study how a predictor can aid in an algorithmic task~\cite{Gupta2017APA, Balcan2021HowMD, Balcan2018DispersionFD, Balcan2018LearningTB, Balcan2018DataDrivenCV, Chawla2019LearningOS}. See the article~\cite{mitzenmacher2020algorithms} for a recent survey.

%% file: icml22/omit3.tex
\section{Omitted Details from \cref{sec:faster_matching}}\label{sec:o3}
\subsection{Proof of \cref{thm:faster-matching}}
% \subsection{Correctness of  \texorpdfstring{\cref{alg:fastmatching}}{Algorithm 2}}
\paragraph{Correctness of \cref{alg:fastmatching}.}
We now prove the correctness of   \cref{alg:fastmatching}.
The following three claims are well known facts about the minimum-weight perfect matching problem (see, for instance~\cite{ahuja1993networkflows}).
\begin{claim}
\label{lem:flow-matching1}
There exists a flow of value $n$ if and only if there exists a perfect matching.
\end{claim}

\begin{claim}
\label{lem:flow-matching2}
If the flow $f$ at the end of the algorithm is a minimum-cost flow of value $n$, then the returned set of edges is a minimum-cost perfect matching.
\end{claim}

\begin{claim}
\label{lem:negative-cycle}
A flow $f$ is a minimum-cost flow if and only if $G_f$ contains no negative cost cycles.
\end{claim}

\begin{lemma}
\label{lem:st-cycle}
The residual graph $G_f$ at the end of   \cref{alg:fastmatching} contains no cycles which include nodes $s$ or $t$.
\end{lemma}

\begin{proof}
At the end of the algorithm, $f$ will have value $n$ and will saturate all edges leaving $s$ as well as all edges entering $t$. Therefore, all edges containing $s$ will be incoming edges and all edges containing $t$ will be outgoing edges. For both nodes, as all of their edges are oriented in the same direction, they cannot be included in any cycles.
\end{proof}

\begin{lemma}
\label{lem:non-negative}
From their initialization in step 10 of   \cref{alg:fastmatching}, all reduced edge costs $c'$ in the residual graph $G_f$ are non-negative.
\end{lemma}

\begin{proof}
Note that any edges containing $s$ or $t$ always have reduced cost 0.
Starting from the costs set at step $10$ of the algorithm, we will prove inductively that for all edges $(i,j)$ in the residual graph s.t. $i, j \notin \{s,t\}$, $c'_{ij} \geq 0$.
From the feasibility of the dual variables $y$, for all $(i,j) \in E$ s.t. $i \in L, j \in R$,
\[
    y_i + y_j \leq c_{ij}.
\]
Therefore, for all left to right edges in the residual graph,
\[
    c'_{ij} = c_{ij} + z_i - z_j = c_{ij} - y_i - y_j \geq 0.    
\]
It remains to consider edges $(j,i) \in E_f$ s.t. $j \in R, i \in L$.
These backwards edges exist if and only if $f_{ij} = 1$.
By step $7$, $f_{ij} = 1$ only if the matching $M$ derived in step $4$ contains $(i,j)$.
This implies that $y_i + y_j = c_{ij}$.
Therefore, for such edges $(j,i) \in E_f$,
\[
    c'_{ji} = -c_{ij} + z_j - z_i = -c_{ij} + y_j + y_i = 0.
\]
This completes the base case: all edges in the residual graph have non-negative reduced cost at step $10$.

For the inductive step, assume going into the while loop that all $c'_e$ for $e=(i,j) \in G_f$ such that $i, j \notin \{s,t\}$ are non-negative.
Consider the new reduced costs for any such edge $(i,j)$ at step 13, noting that $d(s,j) \leq d(s,i) + c'_{ij}$.
Let $c''$ denote the updated costs and $c'$ denote the previous reduced costs:
\begin{align*}
    c''_{ij} 
    = c_{ij} + z_i - z_j &= c'_{ij} + d(s,i) - d(s,j) \\
    &\geq c'_{ij} + d(s,i) - (d(s,i) + c'_{ij}) = 0.
\end{align*}
So, the new costs are also non-negative.

Now, consider the augmentation in step 17 which will change the edges in the residual graph $G_f$.
In particular, some edges $(i,j)$ will disappear and be replaced by edges $(j,i)$ with reduced costs $c'_{ji} = -c'_{ij}$.
As we only augment along edges with reduced costs $c_{ij} = 0$, the reverse edges we create in the residual graph will also have reduced cost 0.
Therefore, the invariant holds that costs remain non-negative, completing the proof.
\end{proof}

\begin{lemma}
\label{lem:progress}
Assume that there exists a perfect matching. If $f$ has flow value less than $n$, then the flow $g$ computed in step 16 of   \cref{alg:fastmatching} will have value at least $1$.
\end{lemma}

\begin{proof}
First, note that the flow $f$ maintained by the algorithm will always have capacity in $\{0,1\}$ as the graph $G$ has unit capacities and we always augment by a maximum flow which will have value either $0$ or $1$ along each edge.
If $f$ has value less than $n$, as the max flow value is $n$ (due to the existence of a perfect matching), there must exist some path of flow value $1$ from $s$ to $t$ in $G_f$.  This implies that $d(s,t)>0$.
Let $P$ be the a shortest path from $s$ to $t$ in $G_f$ as measured by the reduced costs $c'$.
Note that for any edge $(i,j) \in P$, $d(s,j) = d(s,i) + c'_{ij}$ (or else, $P$ is not a shortest path).
So, after updating the reduced costs in step 13, every edge in $P$ must have reduced cost $0$.
Therefore, $P \subseteq E'_f$ and the maximum flow computed in step 16 must have flow value $1$.
\end{proof}

\begin{lemma}
\label{lem:correctness-matching}
\cref{alg:fastmatching} returns a minimum-cost perfect matching, if one exists.
\end{lemma}
\begin{proof}
Assume that a perfect matching exists.
By Lemmas~\ref{lem:flow-matching1} and~\ref{lem:progress}, the algorithm will terminate with a flow of value $n$.
By Lemmas~\ref{lem:flow-matching2} and~\ref{lem:negative-cycle}, it suffices to show that there are no negative cycles in the final residual graph $G_f$.
By~\cref{lem:st-cycle}, any cycles in $G_f$ must only contain edges $(i,j)$ where $i,j \notin \{s,t\}$. 
By~\cref{lem:non-negative}, all reduced costs $c'_{ij}$ of edges $(i,j) \in E_f$ are non-negative.
Let $C \subset E_f$ be any cycle in $G_f$.
The value of $C$ is
\[
    \sum_{(i,j) \in C} c_{ij} = \sum_{(i,j) \in C} c'_{ij} - z_i + z_j \geq \sum_{(i,j) \in C} -z_i + z_j.
\]
As $C$ is a cycle, each vertex incident on $C$ has equally many incoming and outgoing edges, cancelling out the contributions of each $z_u$ so that
\[
    \sum_{(i,j) \in C} -z_i + z_j = 0.
\]
Hence, there are no negative cycles in $G_f$, completing the proof of the correctness of~\cref{alg:fastmatching}.
\end{proof}

% \subsection{Runtime Analysis of \texorpdfstring{\cref{alg:fastmatching}}{Algorithm 2}}
\paragraph{Runtime of \cref{alg:fastmatching}}
We now analyze the runtime of  \cref{alg:fastmatching}. 
\begin{lemma}
\label{lem:firstmatching}
After the first call to maximum cardinality matching in step 4, the partial matching $M$ will contain at least $n - \|y^* - \hat{y}'\|_0$ edges.
\end{lemma}

\begin{proof}
Consider the minimum-weight perfect matching $M^*$ corresponding to the optimal duals $y^*$. Note that the edges in $M^*$ are vertex disjoint since they constitute a matching. Let $M''$ be the set of edges in $M^*$ that are tight under $\hat{y}'$:
\[
    M'' = \{e=ij \in M^*: \hat{y}'_i + \hat{y}'_j = c_e\}.
\]
As $y^*$ is tight for all edges in $M^*$, if the predicted and optimal duals agree on both endpoints of an edge in $M^*$, that edge is also tight under $\hat{y}'$.
Therefore, out of the $n$ edges in $M^*$, at most $\|y^* - \hat{y}'\|_0$ of them do not have tight constraints under $\hat{y'}$.
Equivalently, $|M''| \geq n - \|y^* - \hat{y}'\|_0$.

Consider the call to maximum cardinality matching in step 4.
As $M^*$ is a valid matching and $M'' \subseteq M^*$, $M''$ is a valid matching as well. In addition, all of $M''$'s edges are contained in $E'$ (the set of tight edges under $\hat{y}'$).
Therefore, $M''$ is a valid matching for step 4.
So, the maximum cardinality matching returned by step 4 must have size at least $|M''| \geq n - \|y^* - \hat{y}'\|_0$, completing the proof.
\end{proof}

\begin{lemma}
The total number of iterations of the while loop in step 11 will be at most $\|y^* - \hat{y}'\|_0 $.
\label{lem:whileiters}
\end{lemma}

\begin{proof}
By~\cref{lem:progress}, we will increase the flow value of $f$ by at least $1$ each iteration of the while loop.
By~\cref{lem:firstmatching}, $f$ enters the while loop with value $n - \|y^* - \hat{y}'\|_0$.
Therefore, there can be at most $\|y^* - \hat{y}'\|_0$ iterations.
\end{proof}

\begin{lemma}
The total amount of work done by calls to Ford-Fulkerson in step 16 is $O(m \|y^* - \hat{y}'\|_0)$
\label{lem:fordfulkerson}
\end{lemma}

\begin{proof}
The runtime of Ford-Fulkerson is $O(mf)$ where $f$ is the value of the max flow.
By~\cref{lem:firstmatching}, the flow value can increase by at most $\|y^* - \hat{y}'\|_0$ over all calls to Ford-Fulkerson before we reach a flow of value $n$.
So, Ford-Fulkerson does a total of $O(m \|y^* - \hat{y}'\|_0)$ work, as required.
\end{proof}

We are now ready to prove the main theorem.
\begin{proof}[Proof of \cref{thm:faster-matching}]
By \cref{lem:correctness-matching}, the algorithm returns a minimum-cost perfect matching, if one exists. It remains to prove the runtime.
The first call to matching takes $O(m\sqrt{n})$ time.
Constructing the initial flow and residual graph as well as the corresponding costs takes $O(m)$ time.
By~\cref{lem:whileiters}, there are at most $\|y^* - \hat{y}'\|_0$ iterations of the while loop.
In each iteration, calculating the shortest path distances can be done via Dijkstra's algorithm in $O(m + n\log n)$ time as the reduced costs are always non-negative by~\cref{lem:non-negative}.
Updating the costs and constructing the subgraph $G'_f$ takes $O(m)$ time.
Augmenting along the flow $g$ takes $O(m)$ time.
So, ignoring calls to Ford-Fulkerson, the total running time of work done in the while loop is $O((m+n\log n) \|y^* - \hat{y}'\|_0)$.
By~\cref{lem:fordfulkerson}, all calls to Ford-Fulkerson takes a total of $O(m \|y^* - \hat{y}'\|_0)$.

So, the total runtime of the algorithm is $O(m\sqrt{n} + (m + n\log n) \|y^* - \hat{y}'\|_0)$, as required.
\end{proof}

%% file: icml22/omit-bmatch.tex
\section{Improved Learning-Based Minimum-Weight \texorpdfstring{$b$}{b}-Matching}\label{sec:bmatch}

\begin{algorithm} 
\caption{\label{alg:mwbm-pd}Primal-Dual Scheme for MWBM from~\cite{dinitz2021faster}}
\begin{algorithmic}[1]
\Procedure{MWBM-PD}{$G = (V,E),c,y$}
\State $E' \gets \{ ij \in E \mid y_i + y_j = c_{ij}$ \}
\Comment{Set of tight edges in the dual}
\State $G' \gets (L\cup R \cup \{s,t\}, E' \cup \{ si \mid i \in L\} \cup \{jt \mid j \in R\})$ \Comment{Network of tight edges}
\State $\forall e  \in E(G')$ s.t. $e = si$ or $e= it$, $u_e \gets b_i$ 
\State $u_e \gets \infty$ for all other edges of $G'$
\State $f \gets $ Maximum $s-t$ flow in $G'$ with capacities $u$
\While{Value of $f$ is $< \sum_{i \in L} b_i$}
\State Find a set $S \subseteq L$ such that $\sum_{i \in S} b_i > \sum_{j \in \Gamma(S)} b_j$
% \Comment{Exists by Lemma~\ref{lem:b-matching-augment-set}}
\Comment{Can be found in $O(m+n)$ time}
\State $\epsilon \gets \min_{i \in S,j \in R\setminus \Gamma(S)} \{ c_{ij} - y_i - y_j \}$
\State $\forall i \in S$, $y_i \gets y_i + \eps$
\State $\forall j \in \Gamma(S)$, $y_j \gets y_j - \eps$
\State Update $E',G',u$
\State $f \gets $ Maximum $s-t$ flow in $G'$ with capacities $u$
\EndWhile
\State $x \gets f$ restricted to edges of $G$
\State Return $x$
\EndProcedure
\end{algorithmic}
\end{algorithm}

As a corollary to Theorem~\ref{thm:improved-b-matching}, when combined with the near-linear time rounding procedure from~\cite{dinitz2021faster}, this algorithm gives a fast framework for taking a predicted (possibly infeasible) dual and using it to speed up minimum-weight $b$-matching.

\begin{corollary}
There exists an algorithm which takes as input a (possibly infeasible) integral dual solution $\hat{y}$, produces a feasible dual $\hat{y}'$ s.t.\ $\|\hat{y}' - y^*\|_{b, 1} \leq 5 \|\hat{y} - y^*\|_{b, 1}$, and finds a minimum-weight perfect $b$-matching in $O(mn +  m\|y^* - \hat{y}'\|_{b, 0})$ time, where $y^*$ is an optimal dual solution.
\end{corollary}

The runtime of $O(mn\|y^* - \hat{y}'\|_{b, 1})$ from~\cite{dinitz2021faster} is derived from the fact that each time a maximum flow is found in step 13, the flow value is increased by at least $1$ (due to the integrality of the problem), and each maximum flow can be found in $O(nm)$ time.

To get the improved runtime in~\cref{thm:improved-b-matching}, we can follow essentially the same analysis to that for minimum-weight perfect matching, showing that first call to maximum flow will push a significant amount of flow and then bounding the rest of the work in terms of the remaining flow to be pushed.

\begin{proof}[Proof of \cref{thm:improved-b-matching}]
The correctness of \cref{alg:mwbm-pd} comes from prior work, so it suffices to prove the time complexity.
After the first call to max flow in step 6, the flow value will be at least $\sum_i b_i - \|y^* - \hat{y}'\|_{b, 0}$ by the same argument as~\cref{lem:firstmatching}.
In particular, consider an optimal flow (corresponding to a minimum-weight $b$-matching) $f^*$. Consider the flow $g$ where $g_e = \min\{f^*_e, u'_e\}$ where $u'_e$ is the capacity of edge $e$ in $G'$ in step 6 of the algorithm ($g$ is the subset of $f$ that satisfies the capacities in $G'$). For each edge $e=(u,v)$ where $y^*_u = y'_u$ and $y^*_v = y'_v$, $g_e = f_e$. Conversely, if there is some vertex $u$ where $y^*_u \neq y'_u$, at worst this vertex can invalidate $b_u$ edges as $u$ can be incident on at most $b_u$ edges in $f^*$. So, the value of $g$ is at most $\|y^* - \hat{y}'\|_{b,0}$ less than that of $f^*$. As $f^*$ is optimal, it has value $\sum_i b_i$, and the first call to max flow must push at least $\sum_i b_i - \|y^* - \hat{y}'\|_{b, 0}$ units of flow.

Subsequently, over all calls to maximum flow in step 13, the total amount of flow pushed is at most $\|y^* - \hat{y}'\|_{b, 0}$ and the total number of iterations of the while loop is at most $\|y^* - \hat{y}'\|_{b, 0}$.
By implementing max flow in step 13 by the Ford-Fulkerson algorithm, the total amount of work done in the while loop will be $O(m \|y^* - \hat{y}'\|_{b, 0})$ as Ford-Fulkerson takes linear time per unit of flow and all other work done in the while loop takes linear time per iteration.
As the first call to max flow in step 6 can take time $O(mn)$, we get a total runtime of $O(mn +  m\|y^* - \hat{y}'\|_{b, 0})$, as required.
\end{proof}

%% file: icml22/omit4.tex
\section{Omitted Details from \cref{sec:RE_dual}}\label{sec:o4}
\subsection{Proof of Main Theorem: \cref{thm:RE_dual_rounding}}
 The goal of the section is to prove the correctness and runtime of \cref{alg:round_shortest_paths}. We first need the following auxiliary lemmas, starting from an observation from \cite{goldberg1995scaling}.

\begin{lemma}\label{lem:acyclic}
The graph $G^-$ (after contracting all strongly connected components) is acyclic.
\end{lemma}

\begin{lemma}\label{lem:non_negative}
Consider any edge $e$ such that $\ell_{\hat{y}}(e) \ge 0$ at any stage of \cref{alg:round_shortest_paths}. Then $e$ will always continue to satisfy $\ell_{\hat{y}}(e) \ge 0$.
\end{lemma}
\begin{proof}
Let $e = (u,v)$. We prove the lemma by showing that $\ell_{\hat{y}}(e) \ge 0$ continues to hold after every iteration of the \texttt{while} loop in step $3$. The only way for $\ell_{\hat{y}}(e)$ to change is if one of $y_u$ or $y_v$ is updated in step $9$ of \cref{alg:round_shortest_paths}. If both $u,v \in \cup_{t \ge i^*} L_i$ or both $u,v \not \in \cup_{t \ge i^*} L_i$ then the edge is unchanged. Now if $v \in  \cup_{t \ge i^*} L_i$ but not $u$, then $\ell_{\hat{y}}(e)$ increases by $1$ so $\ell_{\hat{y}}(e) \ge 0$ continues to hold for this iteration. Now suppose that $u \in  \cup_{t \ge i^*} L_i$ but not $v$. In this case, if $\ell_{\hat{y}}(e) $ was strictly greater than zero, i.e., $\ell_{\hat{y}}(e) \ge 1$, then $\ell_u - \ell_v$ only decreases by $1$ so $\ell_{\hat{y}}(e) \ge 0$ is maintained. Lastly we need to consider the possibility that $\ell_{\hat{y}}(e) =0$. In this case, it must be that $v \in  \cup_{t \ge i^*} L_i$ since we can go from $x$ to $v$ via $x \rightarrow u \rightarrow v$ which means $v$ is either in the same layer as $u$ or possibly a higher layer. Both of these scenarios were addressed previously so we are done.
\end{proof}

\begin{lemma}\label{lem:max_layer_num}
At any iteration of the \texttt{while} loop, $i^* \le 2\|\hat{y} - y^*\|_1$ where the quantity $\|\hat{y} - y^*\|_1$ denotes the \textbf{initial} predictor error.
\end{lemma}
\begin{proof}
We first provide a bound for the \emph{very first} iteration of the \texttt{while} loop on Line 3 of \cref{alg:round_shortest_paths}.
Note that $G^{-}$ is acyclic due to Lemma \ref{lem:acyclic}. All paths in $G^{-}$ must use non-negative edges. Any negative edge $e = (u,v)$ has reduced length at least $- (|y^*_u - \hat{y}_u| + |y^*_v - \hat{y}_v|)$. This is because we know that $\ell(u,v) + y^*_u - y^*_v \ge 0$. Thus, the absolute value of the length of $\ell_{\hat{y}}(e)$ is at most 
\[ |\ell_{y^*}(e) - \ell_{\hat{y}}(e)| \le |y^*_u - \hat{y}_u| + |y^*_v - \hat{y}_v|.  \]
Now consider the max $i$ for which $L_i$ exists. This means there is a path $x = u_1, \cdots, u_k = v$ of total length $-i$. We have
\[i = \sum_{j=1}^{k-1} |\ell_{\hat{y}}(u_j, u_{j+1})| \le \sum_{j=1}^{k-1} |y^*_{u_j} - \hat{y}_{u_j}| + |y^*_{u_{j+1}} - \hat{y}_{u_{j+1}}| \le 2\|y^* - \hat{y}\|_1\]
since every vertex $u$ can appear at most twice in the middle summation above.

We now claim that the maximum $i$ is \emph{always} at most $2\|y^* - \hat{y}\|_1$.  \cref{lem:non_negative} implies that non-negative edges (under $\ell_{\hat{y}}$) always stay non-negative. 
Furthermore, the proof of  \cref{lem:non_negative}   tells us that the length of any negative edge is monotonically increasing until the edge becomes non-negative. Therefore any path from $x$ to $v$ in $G^{-}$ in any iteration of the \texttt{while} loop must have also existed in the very first iteration. It follows that most negative distances in $G^-$ are monotonically decreasing every iteration, i.e., becoming less negative. Therefore, the same bound on the number of layers $L_i$ also continues to hold for all instances of $G^-$.
\end{proof}

\begin{proof}[Proof of  \cref{thm:RE_dual_rounding}] The correctness of  \cref{thm:RE_dual_rounding} follows from the fact that the \texttt{while} loop only stops when all reduced edge lengths are non-negative. Therefore, the main challenge is to bound the number of iterations. From the standard analysis of Goldberg's algorithm \cite{goldberg1995scaling}, we know that each iteration of the \texttt{while} loop takes $O(m)$ time. This is because $G^{-}$ is an acyclic graph and thus, finding the layers $L_i$ and all subsequent computations can be done in $O(m)$ time. Thus, it remains to bound the number of \texttt{while} loop iterations. 

Now call a vertex $v$ touched if $v \in L_{i^*}$ for $i^*$ defined in step $8$ of \cref{alg:round_shortest_paths}. Note that for a vertex to be touched, it must exist in some layer and therefore has a negative incoming edge. We now claim that every time a vertex is touched, its most negative \emph{incoming} edge increases in length by $+1$. 

Indeed, let $(u,v)$ be the most negative incoming edge to $v$ and suppose that $v \in L_{i*}$. Vertex $u$ cannot exist in layer  $L_t$ for some $t > i^*$ since we can consider the path $x \rightarrow u \rightarrow v$ which implies $v$ must exist in a larger layer than $u$. Thus when $v$ is touched, the edge length $(u,v)$ must increase by $1$. From  \cref{lem:max_layer_num}, we know that layer $L_{i*}$ has at least $n/(2\|\hat{y} - y^* \|_1$) many vertices since there are at most $2 \|\hat{y} - y^* \|_1$ layers which partition all $n$ vertices. Therefore, at least $n/(2\|\hat{y} - y^* \|_1)$ many vertices get touched in every iteration of the \texttt{while} loop. Each vertex can only get touched at most $O(\|y^* - \hat{y} \|_{\infty})$ times since the most negative edge length in the very beginning of \cref{alg:round_shortest_paths} has absolute value at most $O(\|y^* - \hat{y} \|_{\infty})$. This implies that the number of \texttt{while} loop iterations is at most $O(\|\hat{y} - y^* \|_1 \cdot \|y^* - \hat{y} \|_{\infty})$. Since every \texttt{while} loop iteration takes $O(m)$ time, the bound of $O(m \|\hat{y} - y^* \|_1 \cdot \|y^* - \hat{y} \|_{\infty})$ follows. Note that we could have also used Goldberg's algorithm after getting the reduced edge lengths from $\ell_{\hat{y}}$ with no further modifications to get time $O(m\sqrt{n} \log(\|y^* - \hat{y} \|_{\infty})$. Therefore, running these two algorithms in parallel implies the claimed running time.
\end{proof}

\subsection{All-Pair Shortest Paths}\label{sec:apsp}
We observe that \cref{thm:RE_dual_rounding} implies the following runtime for finding all pairs shortest paths on a graph.

\begin{theorem}\label{thm:faster_shortest_paths_dual}
There exists an algorithm which takes as input predicted reduced edge duals $\hat{y} : V \rightarrow \mathbb{Z}$ and outputs all pair shortest paths in $O(m \min\{\|\hat{y}-y^*\|_1 \cdot \|\hat{y}-y^*\|_{\infty}, \sqrt{n} \log(\|\hat{y}-y^*\|_{\infty}))\} + mn + n^2 \log n)$ time where $y^*: V \rightarrow \mathbb{Z}$ denotes a feasible set of reduced edge length duals.
\end{theorem}
\begin{proof}
Consider \cref{alg:round_shortest_paths}. It applies \cref{alg:round_shortest_paths} to round $\hat{y}$ into a feasible RE dual $\hat{y}'$. Then we can run Dijkstra's algorithm starting from all vertices in time $O(mn + n^2 \log n)$. The running time follows from  \cref{thm:RE_dual_rounding}.
\end{proof}

\begin{algorithm}[H]
\caption{\label{alg:round_shortest_paths-2} Learning-based Shortest Paths}
\begin{algorithmic}[1]
\State \textbf{Input:} Graph $G = (V,E)$, predicted duals $\hat{y}: V \rightarrow \mathbb{Z}$
\Procedure{Faster-Shortest-Paths}{$G, \hat{y}$}
\State $\hat{y}' \gets $\text{Round-RE-Duals}($G, \hat{y}$)
\Comment{$\hat{y}'$ is a feasible RE Dual}
\For{all $v \in V$}
\State Run Dijkstra's algorithm starting from $v$ 
\EndFor
\State Return all shortest paths found from all vertices
\EndProcedure
\end{algorithmic}
\end{algorithm}

%% file: icml22/additional-reduction.tex
\section{Additional Reductions for Learning-Based Graph Algorithms}\label{sec:more-red}
\subsection{Degree-Constrained Subgraph from Matching}
The degree constrained subgraph (DCS) problem is defined as follows.
We are given an undirected multigraph $G=(V,E)$ (we will only be considering bipartite graphs) as well as a set of desired upper and lower bound on each vertex's degree: $l_i \leq d_i \leq u_i$ for all $i \in V$.
A DCS is an edge-induced subgraph of $G$ where the degree conditions are satisfied.
A DCS is called \emph{complete} if each degree achieves its upper bound: $d_i = u_i$ for all $i \in V$.

The maximum perfect DCS and maximum weights DCS problems correspond to maximum perfect matching and maximum weight matching, respectively. Note that the DCS versions of these problems generalize the matching versions by setting $l_i = 0$ and $u_i = 1$ for all vertices. Next, we will show that DCS can also be reduced to matching following the reduction given in \cite{gabow1985scaling}.

First consider the maximum perfect DCS problem. Let $G=(L, R,E)$ be the corresponding multigraph with degree bounds $l_i, u_i$ for $i \in V$.
We will build a corresponding bipartite graph $H = (L', R', E')$ as follows.
\begin{itemize}
    \item For each vertex $i$ in $G$, create a complete bipartite graph $K_{\delta, d}$ where $d=d_i$ is the degree of $i$ in $G$ and $\delta = d_i - u_i$ is how many edges need to be removed from $i$ to meet the upper bound.
    We will call the $\delta$ side of $K_{\delta, d}$ \emph{internal} nodes and the $d$ side \emph{external nodes}.
    Without loss of generality, assume $i \in L$. Then, the external side of $K_{\delta, d}$ is in $L'$ and the internal side is in $R'$.
    
    \item Associate each of $i$'s edges in $G$ with one of its external nodes in $H$. Specifically, for each $(i,j) \in E$, there will be an edge between one of $i$'s and one of $j$'s external nodes in $H$ and both of those nodes will not be neighbors with any other external nodes.
    Note that as $G$ is bipartite, with these added edges, $H$ will still be bipartite.
    
    \item For each of these external-external edges, give them costs in $H$ corresponding to their costs in $G$.
\end{itemize}

First, note that a perfect matching in $H$ corresponds to a perfect DCS in $G$. For each node $i$ in $G$, all of its $\delta$ internal nodes in $H$ will be matched, meaning that exactly $u_i$ of its external nodes are matched with other external nodes. As each of these external-external edges correspond to edges in the original edgeset $E$, this means that $i$ will have degree $u_i$ in the subgraph induced by the external-external edges in the perfect matching, as required.

Similarly, it is easy to see that every perfect DCS in $G$ corresponds to a perfect matching in $H$, so optimizing over perfect matchings/DCS's are equivalent, completing the reduction.

Assume that $G$ had $n$ vertices and $m$ edges (counting copies). In $H$, we will have $O(m)$ total vertices and $O(m \cdot d_{max})$ total edges where $d_{max}$ is the maximum degree of any vertex in $G$. \cref{alg:reduction} gives us the following corollary.

\begin{theorem}\label{thm:dcs_reduction}
Given a maximum weight perfect DCS problem on input graph $G=(V,E)$ with $n$ vertices, $m$ edges, and maximum degree $d_{max}$, there exists an algorithm which takes takes as input a predicted dual solution $\hat{y}$ to an instance of maximum weight perfect matching derived from $G$, near-optimally rounds the dual to a feasible solution $\hat{y}'$, and solves the DCS in time $O(m^{3/2} d_{max} + (m d_{max} + m\log m) ||y^* - \hat{y}'||_0)$.
\end{theorem}

\subsection{Minimum-Cost 0-1 Flow from Degree-Constrained Subgraph}
The reduction bares resemblance to the reduction from shortest path from matching and is also due to~\cite{gabow1985scaling}.
We are given a directed graph $G$ with unit capacities and integral edge costs $a_{ij}$.
We want to find a minimum cost flow of flow value $v$.
We will construct a bipartite multigraph $H=(L,R,E)$ for the DCS problem as follows.
\begin{itemize}
    \item For each vertex $i \in G$, make two copies $i_1 \in L$ and $i_2 \in R$.
    \item Add $mindegree(i)$ copies of the edge $(i_1, i_2)$ to $H$ each with weight $0$. Where $mindegree(i)$ is the minimum of $i$'s indegree and outdegree.
    \item For each edge $(i,j)$ in $G$, add an edge $(j_1, k_2)$ to $H$ with weight $-a_{jk}$.
    \item Set the degree constrains $u_{i_1} = u_{i_2} = mindegree(i)$ for all $i \neq s,t$.
    Set $u_{s_2} = mindegree(s)$, $u_{s_1} = u_{s_2} + v$,  $u_{t_2} = mindegree(t)$, $u_{t_1} = u_{t_2} + v$.
\end{itemize}

Note that the number of vertices and edges in $H$ are at most twice those in $G$.

\begin{theorem}\label{thm:01flow_reduction}
Given a minimum-cost 0-1 flow problem on input graph $G=(V,E)$ with $n$ vertices, $m$ edges, and maximum degree $d_{max}$, there exists an algorithm which takes takes as input a predicted dual solution $\hat{y}$ to an instance of maximum weight perfect matching derived from $G$, near-optimally rounds the dual to a feasible solution $\hat{y}'$, and solves the DCS in time $O(m^{3/2} d_{max} + (m d_{max} + m\log m) ||y^* - \hat{y}'||_0)$.
\end{theorem}

% \paragraph{Minimum-Weight Edge Cover from Matching.} 
% Consider a weighted graph $G$. A subset of edges $C \subseteq E$ forms an edge cover of $G$ iff  every vertex of $G$ is incident to at least one edge in $C$. In the min-weight edge cover problem, we wish to find an edge cover of minimum weight. There is an efficient reduction from minimum-weight edge cover to minimum-weight perfect matching on bipartite graphs. We outline the reduction following

% The reduction creates a graph $H$ as follows. $H$ will be the union of $G$ and $G_1$ where $G_1$ is an identical copy of $G$ with different edge weights. First, set all weights of ...

% The reduction is given in Section 8.2.9 here \url{https://dl.acm.org/doi/pdf/10.5555/1882757.1882765}. If all starting edge-weights are non-negative, then we can also do maximum-weight perfect matching, see \url{https://cstheory.stackexchange.com/questions/14690/reducing-a-minimum-cost-edge-cover-problem-to-minimum-cost-weighted-bipartie-per/14905#14905}.

% \paragraph{Other Matching Variants and Matching} Some matching variants, like some formulations of `fair' matching, also reduce to maximum weight matching; see \url{https://www.sciencedirect.com/science/article/pii/S030439750700624X}

\subsection{Diameter to Shortest Paths} The diameter of a graph is defined as the largest distance between any pair of vertices. All exact algorithms for calculating the diameter on general weighted graphs all rely on computing all pairs shortest paths (and there is evidence that this approach is unavoidable \cite{DalirrooyfardW21}). Our learning-augmented algorithm for computing shortest-paths of \cref{sec:RE_dual} gives us the following corollary for computing the diameter of an input graph which follows by first rounding to a valid reduced edge length dual of \cref{def:reduced_edge} and running all pairs shortest paths using Dijkstra's algorithm on the resulting graph with non-negative weights.

\begin{theorem}\label{thm:diameter_reduction}
Given an input graph $G$ with $n$ vertices and $m$ edges with possibly negative integer edge lengths given by $\ell$, there exists an algorithm which takes takes as input a predicted dual solution $\hat{y}$ to the reduced edge length dual on $G$ and computes the diameter of $G$ in time 
\[O( m \min\{\|\hat{y}-y^*\|_1 \cdot \|\hat{y}-y^*\|_{\infty}, \sqrt{n} \log(\|\hat{y}-y^*\|_{\infty})\} )+ \tilde{O}(mn).\]
\end{theorem}

\begin{remark}
Note that the an algorithm which doesn't use any learned predictions for computing shortest paths on a graph with negative weights, such as the Bellman-Ford algorithm, would have taken time $O(mn^2)$ to compute the diameter.
Note that we could have also reduced the diameter problem to matching by using the reduction from shortest paths to matching. However the reduction used in \cref{thm:diameter_reduction} is simpler as we don't need to compute any new graphs.
\end{remark}

%% file: icml22/omit6.tex
\section{Omitted Details from Section 6}\label{sec:o6}
Our results generally follow from bounding the pseudo-dimension of the loss function and applying standard uniform convergence for PAC learning. 

\begin{definition}[pseudo-dimension] 
Let $\mathcal{F}$ be a class of functions $f: X \rightarrow \mathbb{R} .$ Let $S=\left\{x_{1}, x_{2}, \ldots, x_{s}\right\} \subset X$ We say that that $S$ is shattered by $\mathcal{F}$ if there exist real numbers $r_{1}, \ldots, r_{s}$ so that for all $S^{\prime} \subseteq S$, there is a function $f \in \mathcal{F}$ such that $f\left(x_{i}\right) \leq r_{i}$ if and only if $ x_{i} \in S^{\prime}$ for all $i \in[s]$. The pseudo-dimension of $\mathcal{F}$ is the largest $s$ such that there exists an $S \subseteq X$ with $|S|=s$ that is shattered by $\mathcal{F}$.
\end{definition}
For a class of loss functions with bounded range and pseudo-dimension, the following lemma provides a PAC learning guarantee. 
\begin{lemma}[uniform convergence; e.g., \cite{anthony1999neural}]\label{lem:uni-dim}
 Let $\mathcal{D}$ be a distribution over a domain $X$ and $\mathcal{F}$ be a class of functions $f: X \rightarrow[0, H]$ with pseudo-dimension $d_{\mathcal{F}}$. Consider $s$ i.i.d.\ samples $x_{1}, x_{2}, \ldots, x_{s}$ from $\mathcal{D} .$ There is a universal constant $c_{0}$, such that for any $\epsilon>0$ and $p \in(0,1)$, if $s \geq c_{0}\left(\frac{H}{\epsilon}\right)^{2}\left(d_{\mathcal{F}}+\ln (1 / \delta)\right),$ then we have
$$
\left|\frac{1}{s} \sum_{i=1}^{s} f\left(x_{i}\right)-\mathbb{E}_{x \sim \mathcal{D}}f(x)\right| \leq \epsilon
$$
for all $f \in \mathcal{F}$ with probability at least $1-\delta .$
\end{lemma}

\subsection{Proof of \cref{thm:gen-learn-uni}}
To prove \autoref{thm:gen-learn-uni}
Let $f_{1,h} (c) = \ell_1 (h,c ) $ for $h \in \mathcal{H}$.
The following  lemma provides a bound on the pseudo-dimension of the family $\mathcal{F}_1 = \{ f_{1,h} (c)  : h \in \mathcal{H} \}$. 
\begin{lemma}[\cite{dinitz2021faster}]\label{lem:pseud-dim}
 The pseudo-dimension of $\mathcal{F}_1$ is bounded above by $O(d\log d)$.
\end{lemma}
Now we are ready to prove our learnability result \cref{thm:gen-learn-uni}.
\begin{proof}[Proof of \cref{thm:gen-learn-uni}]
Given $s$ samples $c_1,c_2\cdots, c_s$, the algorithm performs empirical risk minimization on the loss $\hat \ell(h)= \sum_{i=1}^s \|h^*(c_i) - h\|_1$. The algorithm runs in polynomial time by the efficient optimization assumption.

Moreover, since $\mathcal{H} \subseteq \mathbb{R}^d$ and $\mathcal{H}$   has bounded range, we have that any function in $\mathcal{F}_1$ is bounded by $dM$. Therefore, the sample   complexity and error bound follows from \cref{lem:uni-dim} and \cref{lem:pseud-dim}.   
\end{proof}

\subsection{Proof of \cref{thm:gen-learn-uni-infty}}

\begin{proof}[Proof of \cref{thm:gen-learn-uni-infty}]
Similar to the $\ell_1$ learnability theorem, the algorithm simply finds the empirical minimizer of $\hat \ell(h)= \sum_{i=1}^s \|h^*(c_i) - h\|_\infty$. The algorithm runs in polynomial time by the efficient optimization assumption. 

Let $f_{\infty,h} (c) = \ell_\infty (h,c ) $ for $h \in \mathcal{H}$.
It now suffices to bound the pseudo-dimension of the family $\mathcal{F_\infty} = \{ f_{\infty,h} (c)  : h \in \mathcal{H} \}$ and then apply the uniform convergence lemma (\cref{lem:uni-dim}). Now observe that the pseudo-dimension of $\mathcal{F}_\infty$ can be in turn bounded by the VC dimension of axis-aligned hyperrectangles in $\mathbb{R}^d$, which is known to be $2d$ \cite{mohri2018foundations}.
\end{proof}

\subsection{Details on Learnability Via Arithmetic Complexity}\label{sec:arith}
Suppose we have any loss function $L(h,G) \in   \R$ which represents how well a hint vector performs on some input $G$. 
 %An example of $L$ could be squared error from the output of $h$ to some true set of edge values for edges in $G$. Another choice of $L$ could be the $\ell_1$ loss function studied in Section \ref{sec:learnability_pd}. 
 For notational simplicity, we define $\mathcal{A}$ as the class of functions in $h$ composed with $L$:
\[\mathcal{A} := \{L \circ h : h \in \mathcal{H} \}.  \]
We also assume that the range of $L$ is equal to $[0,H]$ and that all graphs $G$ can be represented as a feature vector in $\R^m$. 

Again, we aim to learn the best function $h \in \mathcal{H}$ which minimizes the following objective:
\begin{equation}\label{eq:best_h}
    \E_{c \sim D}[L(h,G)].
\end{equation}
Towards this end, we let $h^*$ be such the optimal $h \in \mathcal{H}$. We also assume that for each instance $G$ and each $h \in \mathcal{H}$, $L \circ h(G)$ can be computed in time $T(m,d)$. For example, suppose graphs drawn from $\mathcal{D}$ possess edge features in $\R^d$ for some $d$ and our family $\mathcal{H}$ is parameterized by a single vector $\theta \in \R^d$ and represents linear functions which report the dot product of each edge feature with $\theta$. Then it is clear that $T(m,d)$ is a (small) polynomial in the relevant parameters.

The result of this section is to bound the pseudo-dimension of $\mathcal{A}$. After obtaining a bound, we can readily apply   \cref{lem:uni-dim} as we did in the proof of  \cref{thm:gen-learn-uni} in   \cref{sec:learnability_pd}.

\begin{theorem}[Learnability via  computational complexity]\label{lem:psudeo_dim_alt}
Suppose that any $a \in \mathcal{A}$ takes $T(m,d)$ time to compute given any graph $H$ drawn from $\mathcal{D}$. Then the pseudo-dimension of $\mathcal{A}$ is $O(\text{poly}(T(m,d))$.
\end{theorem}

To prove \cref{lem:psudeo_dim_alt}, we first relate the pseudo-dimension to the VC dimension of a related class of threshold functions. This relationship has been fruitful in obtaining learning bounds in a variety of works such as \cite{feldman_gaussians, barycenter}.

\begin{lemma}[Pseudo-dimension to VC dimension, Lemma $10$ in \cite{feldman_gaussians}]\label{lem:PD_to_VC}
For any $a \in \mathcal{A}$, let $B_a$ be the indicator function of the region on or below the graph of $a$, i.e., $B_a(x,y)  =  \text{sgn}(a(x)-y)$. The pseudo-dimension of $\mathcal{A}$ is equivalent to the VC-dimension of the subgraph class $B_{\mathcal{A}}=\{B_a \mid a \in \mathcal{A}\}$.
\end{lemma}

The following theorem then relates the VC dimension of a given function class to its computational complexity, i.e., the complexity of computing a function in the class in terms of the number of operations needed.

\begin{lemma}[Theorem $8.14$ in \cite{anthony1999neural}]\label{lem:VC_bound}
Let $w: \R^{\alpha} \times \R^{\beta} \rightarrow \{ 0,1\}$, determining the class
\[ \mathcal{W} = \{x \rightarrow w(\theta, x) : \theta \in \R^{\alpha}\}. \]
Suppose that any $w$ can be computed by an algorithm that takes as input the pair $(\theta, x) \in \R^{\alpha} \times \R^{\beta}$ and returns $w(\theta, x)$ after no more than $r$ of the following operations:
\begin{itemize}
    \item arithmetic operations $+, -,\times,$ and $/$ on real numbers,
    \item jumps conditioned on $>, \ge ,<, \le ,=,$ and $=$ comparisons of real numbers, and
    \item output $0,1$,
\end{itemize}
then the VC dimension of $\mathcal{W}$ is $O(\alpha^2 r^2 + r^2 \alpha \log \alpha)$.
\end{lemma}

Combining the previous results allows us prove  \cref{lem:psudeo_dim_alt}. At a high level, we are instantiating  \cref{lem:VC_bound} with the complexity of \textit{computing} any function in the function class $\mathcal{A}$.